\documentclass[lettersize,journal]{IEEEtran}
\usepackage{amsmath,amsfonts}
\usepackage{algorithmic}
\usepackage{algorithm}
\usepackage{array}
\usepackage{subcaption}
\usepackage{textcomp}

\usepackage{dirtytalk}
\usepackage{stfloats}
\usepackage{url}
\usepackage{verbatim}
\usepackage{graphicx}
\usepackage{cite}
\usepackage{wrapfig}
\usepackage{url}            
\usepackage{booktabs}       
\usepackage{amssymb}
\usepackage{nicefrac}       
\usepackage{microtype}      
\usepackage{tikz}
\usepackage{soul}
\usepackage{mathtools}
\usepackage{multicol}

\usepackage[linesnumbered,ruled,vlined,algo2e,algoruled,boxed]{algorithm2e}
\SetKwInput{KwInit}{Initialize}

\newtheorem{definition}{Definition}
\newtheorem{lemma}{Lemma}
\newtheorem{remark}{Remark}
\newtheorem{theorem}{Theorem}

\newcommand{\bL}{{\mathbf L}}

\usepackage{etoolbox}
\makeatletter
\patchcmd{\@begintheorem}{\textit}{\textbf}{}{}
\makeatother

\usepackage[hidelinks]{hyperref}
\hypersetup{
    colorlinks=true,
    linkcolor=blue,
    filecolor=magenta,
    urlcolor=cyan,
    citecolor=red
}

\def\matt#1{\begin{bmatrix}#1\end{bmatrix}}

\begin{document}
\title{A Control-Barrier-Function-Based Algorithm for Policy Adaptation in Reinforcement Learning}

\author{Wenjian Hao, Zehui Lu, Nicolas Miguel, Shaoshuai Mou
\thanks{The authors are with the School of Aeronautics and Astronautics, Purdue University, West Lafayette, IN 47907, USA. {\tt\small \{hao93,  lu846, nmiguel, mous\}@purdue.edu}.}

\thanks{The experimental video is available at \url{https://youtu.be/iYwDWh-JNkU}.}
}







\maketitle

\begin{abstract}
This paper considers the problem of adapting a predesigned policy, represented by a parameterized function class, from a solution that minimizes a given original cost function to a trade-off solution between minimizing the original objective and an additional cost function. The problem is formulated as a constrained optimization problem, where deviations from the optimal value of the original cost are explicitly constrained. To solve it, we develop a closed-loop system that governs the evolution of the policy parameters, with a closed-loop controller designed to adjust the additional cost gradient to ensure the satisfaction of the constraint. The resulting closed-loop system, termed control–barrier–function–based policy adaptation, exploits the set-invariance property of control barrier functions to guarantee constraint satisfaction. The effectiveness of the proposed method is demonstrated through numerical experiments on the Cartpole and Lunar Lander benchmarks from OpenAI Gym, as well as a quadruped robot, thereby illustrating both its practicality and potential for real-world policy adaptation.
\end{abstract}


\section{Introduction}
\IEEEPARstart{R}{einforcement} learning (RL) has recently gained significant attention in robotics due to the growing complexity of robotic systems \cite{mahadevan1992automatic,zhu2019dexterous,hwangbo2019learning,selim2022safe} and the challenges posed by robotic environments. For instance, interactions in manipulation and locomotion tasks are difficult to predict, and the intricate dynamics of contact and friction forces are challenging to model accurately \cite{agarwal2023legged}. To address these challenges, RL offers a model-free way for learning optimal policies that maximize cumulative rewards (or equivalently minimize cumulative stage costs) over time, without requiring explicit knowledge of the system dynamics.

In standard RL, robots learn policies through trial-and-error interactions with the environment, commonly implemented using an actor-critic framework \cite{sutton1992reinforcement}, where the critic evaluates policy performance and the actor updates the policy parameters accordingly. However, training optimal policies from scratch in high-dimensional, nonlinear robotic systems often demands a large number of training episodes. Moreover, when task specifications change or new objectives are introduced, retraining from scratch may significantly increase both time and computational costs. These challenges highlight the importance of efficient policy adaptation methods that enable robots to achieve new objectives while leveraging existing pretrained policies.
\begin{figure}
\centering
\begin{subfigure}{0.80\linewidth}
\centering
\includegraphics[width=\linewidth]{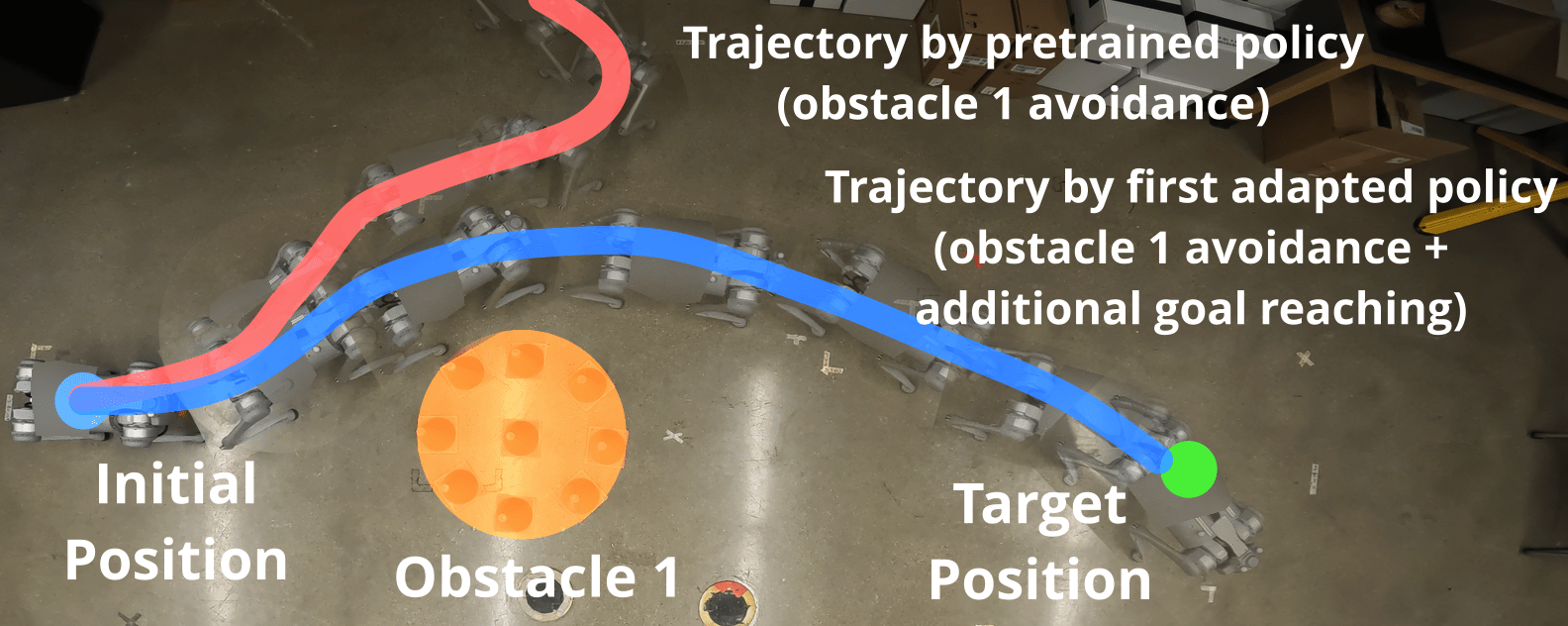}
\caption{Pretrained policy: obstacle 1 avoidance (red trajectory). Adapted policy 1: obstacle 1 avoidance + goal reaching (blue trajectory).}
\label{fig:graphic_abs:1}
\end{subfigure}
\begin{subfigure}{0.80\linewidth}
\centering
\includegraphics[width=\linewidth]{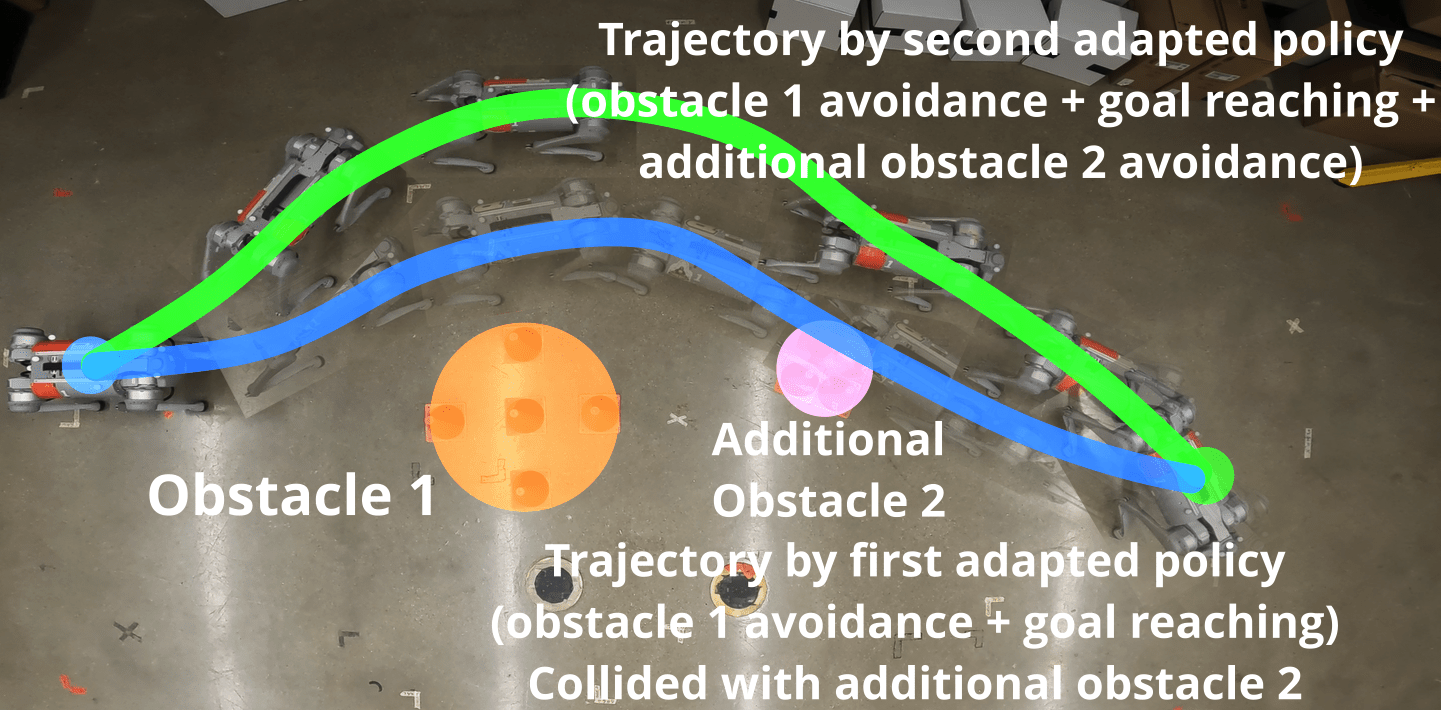}
\caption{Pretrained policy: equivalently adapted policy 1, obstacle 1 avoidance + goal reaching (blue trajectory). Adapted policy 2: obstacle 1 avoidance + goal reaching + additional obstacle 2 avoidance (green trajectory).}
\label{fig:graphic_abs:2}
\end{subfigure}
\caption{A robot executes pretrained policies to perform original tasks. Adapted policies enable the robot to fulfill both original tasks and additional tasks.}
\label{fig:graphic_abs}
\end{figure}

Control barrier functions (CBFs) were introduced in response to the growing need for feasible safety-critical controller design \cite{ames2019control}. Traditional safety-critical optimal control (OC) methods either fail to account for system dynamics feasibility or incorporate safety via potentially nonlinear constraints. These methods can lead to computational intractability and suffer from the curse of dimensionality \cite{xiao2023safetysurvey}. To address these issues, CBFs provide safety guarantees while maintaining computational tractability, leveraging Nagumo’s principle of invariance \cite{ames2019control}. The application of CBFs in existing research primarily focuses on collision avoidance, where the set of states leading to collisions is defined as the unsafe set, which is then treated as a constraint within a CBF quadratic programming (CBF-QP) problem. In those frameworks, the CBF-QP minimally adjusts the reference control input to avoid entering the unsafe set. Consequently, the CBF-QP is typically considered an additional control to a reference closed-loop controller, providing corrections to the reference input when safety risks are detected. While CBFs have become well-established in collision avoidance contexts, their use for optimality guarantees is less explored. Notably, recent work \cite{allibhoy2023} explores the use of CBFs to ensure optimality, utilizing them to design safe gradient flows for solving constrained nonlinear optimization problems. 

Motivated by the above challenges of policy adaptation and the promising capabilities of CBFs, this paper presents a closed-loop system that governs the dynamics of policy parameters. The proposed method minimizes an additional objective while ensuring near-optimal performance with respect to the original task throughout the learning process. The main contributions are summarized as follows: \begin{itemize}
    \item We formulate policy adaptation in RL as a constrained optimization problem over the parameters of a predesigned optimal policy, where the additional objective is minimized subject to constraints on the deviation from the original optimal cost.
    \item We propose a closed-loop system for solving the constrained problem, in which a closed-loop controller modifies the gradient of the additional objective to ensure bounded deviation while improving efficiency by avoiding repeated solutions of the original optimization problem.
\end{itemize}

\subsection{Related Work}\label{relatedwork}
This subsection provides an overview of related work on policy adaptation, focusing on the areas of
tuning OC systems and transfer learning in RL.
\subsubsection{Tuning OC Systems}
RL and OC both aim to design closed-loop controllers that minimize an objective function \cite{powell2012ai, kober2013reinforcement}. Each relies on a system defined by states and inputs, with a model (plant or environment) describing state transitions. Tuning OC systems means adjusting parameters within the OC framework to minimize an additional loss function and meet extra performance requirements.

Early work in the OC community approached this through neighboring extremal optimal control (NEOC) \cite{bryson1975applied, rai2023closed}. NEOC studies how small variations in control preserve optimality when the initial condition of a nonlinear system changes slightly. This is achieved via second-variation theory, which reduces the problem to solving a time-varying linear-quadratic control problem along the original optimal trajectory.
Subsequent research examined perturbations in other factors, such as changes in parameters within the loss function. For example, \cite{rehbock1992computational} addressed time-varying parameters in nonlinear OC and proposed an algorithm to compute gradients of the optimal index with respect to these parameters, reusing pre-computed solutions. Later work, such as \cite{jiang2015optimal}, studied how larger parameter shifts affect nonlinear OC. Since most of these methods were continuous-time, extending NEOC to discrete-time systems became a natural step \cite{ghaemi2009neighboring}.

Recently, Pontryagin's Differentiable Programming (PDP) \cite{jin2020pontryagin} differentiates Pontryagin’s Maximum Principle (PMP), allowing parameters to appear in controls, objectives, and dynamics.
Building on PDP, a cooperative tuning method is proposed for multi-agent systems, where optimal trajectories are tuned in a distributed fashion \cite{9992969}.
Although NEOC and PMP-based approaches are effective, they require knowledge of the full optimal trajectory—a significant limitation in robotics, where tasks often have infinite horizons or high complexity. They also involve solving extra linear-quadratic problems, which become intractable in high-dimensional systems. To overcome these challenges, an online control-informed learning framework was recently introduced \cite{liang2025online}. Unlike trajectory-based methods, it incrementally tunes OC systems using single data points in a recursive fashion.

\subsubsection{Transfer Learning}
Transfer Learning (TL), originally introduced in educational psychology \cite{skinner1965science}, is a machine learning paradigm where knowledge from one domain or task improves performance in another. In RL, TL accelerates learning in new but related tasks by leveraging prior experience \cite{taylor2009transfer}, and knowledge can also be transferred through expert action distributions \cite{czarnecki2019distilling}.

Methodologically, TL is commonly categorized into reward shaping, inter-task mapping, representation transfer, and policy transfer \cite{zhu2020transfer}. Reward shaping adjusts the reward distribution in the target domain using external knowledge to guide policy learning \cite{1999Ngpolicy_invariance,harutyunyan2015expressing}. A related line of work incrementally adapts reward functions based on human feedback, which can also be viewed as interactive reward shaping \cite{9852712}. Inter-task mapping focuses on mapping state–action spaces between source and target domains \cite{taylor2007transfer,gupta2017learning}, while representation transfer leverages shared feature representations such as value functions \cite{devin2017learning} or Q-functions \cite{rusu2016progressive}.

Of these, policy transfer is most relevant to this paper. Here, knowledge is encoded in pretrained source-task policies and transferred to the target task. Policy distillation is one approach, where multiple expert policies are distilled into a student policy by minimizing divergence in action distributions \cite{ross2011reduction,yin2017knowledge,czarnecki2019distilling}. Another is policy reuse, which directly incorporates source policies into target learning \cite{barreto2017successor}. Policy transfer is closely related to imitation learning, as it discourages deviations from the source policy. However, the transferred policy may not always replicate source behavior during training or deployment.

While TL methods primarily transfer knowledge from source tasks, this paper addresses policy adaptation in RL by leveraging the \textit{invariant set} property of control barrier functions (CBFs), offering a conceptually distinct approach.

The rest of this paper is organized as follows. Section \ref{sec:problem} formulates the problem of interest. Section \ref{sec:main} presents the proposed algorithm and its associated theoretical guarantees. Section \ref{sec:sim_sim} discusses the optimization strategy and the selection of key parameters, and further demonstrates the application of the method to policy adaptation in RL through simulation studies. Section \ref{exp_dog} describes hardware experiments that validate the effectiveness of the approach. Section \ref{Conc} concludes the paper and discusses limitations and future directions. 

\textbf{Notations.} 
$\lVert \cdot \lVert$ denotes the Euclidean norm. Given a matrix $A\in\mathbb{R}^{n\times m}$, $A'$ denotes its transpose.

\section{The Problem}\label{sec:problem}
This section first formulates the problem under consideration and subsequently interprets it as a policy adaptation problem within the reinforcement learning (RL) framework.

\subsection{Problem Formulation}
Let $\boldsymbol{\theta}\in\mathbb{R}^p$ denote the vector of tunable parameters. Define $G(\boldsymbol{\theta}):\mathbb{R}^n\times\mathbb{R}^m\rightarrow\mathbb{R}$ and $J(\boldsymbol{\theta}):\mathbb{R}^n\times\mathbb{R}^m\rightarrow\mathbb{R}$ as the original and additional objective functions, respectively, both assumed to be Lipschitz continuous. Let $\boldsymbol{\theta}_G^*\in\mathbb{R}^p$ denote a given minimizer of $G(\boldsymbol{\theta})$, i.e., \begin{equation}
    \boldsymbol{\theta}_G^* = \arg\min_{\boldsymbol{\theta}\in\mathbb{R}^p} G(\boldsymbol{\theta}). \nonumber
\end{equation} The \textbf{problem of interest} is to find the optimal parameters of $J(\boldsymbol{\theta})$ within a bounded neighborhood near the known reference $G(\boldsymbol{\theta}_G^*)$. Specifically, this neighborhood is formalized as the following constraint:
\begin{equation}\label{eq_presol_cons}
G(\boldsymbol{\theta}_G^*) - G(\boldsymbol{\theta}) + c \geq 0,
\end{equation}
where $c\geq 0$ is a relaxation variable, which measures the tolerance for deviation from $G(\boldsymbol{\theta}_G^*)$. It is introduced to address cases where the optimal solution sets of $J$ and $\boldsymbol{\theta}_G^*$ do not intersect. By allowing a bounded relaxation with respect to $G(\boldsymbol{\theta}_G^*)$, the inclusion of $c$ ensures feasible solutions that balance performance across conflicting objectives. 

Accordingly, this paper aims to solve the following constrained optimization problem:
\begin{equation}\label{eq_opti_problem}
    \begin{aligned}
\min_{(\boldsymbol{\theta},c)\in\mathbb{R}^{p+1}} &J(\boldsymbol{\theta}) + wc^2\\
\text{subject to} \quad &\underbrace{G(\boldsymbol{\theta}_G^*) - G(\boldsymbol{\theta})}_{h(\boldsymbol{\theta})} + c\geq 0, \ c \geq 0, \\
    \end{aligned}
\end{equation}
where $w\geq 0$ denotes the constant penalty weight for $c$. 

Let $(\boldsymbol{\theta}^*,c^*)$ denote the optimal solution of \eqref{eq_opti_problem}. Problem \eqref{eq_opti_problem} can be interpreted as a policy adaptation problem in RL. When $c^*=0$, policy adaptation corresponds to finding $\boldsymbol{\theta}^*$ that jointly minimizes both $J(\boldsymbol{\theta})$ and $G(\boldsymbol{\theta})$ while starting from $\boldsymbol{\theta}_G^*$. When $c^*>0$, the solution $\boldsymbol{\theta}^*$ represents a trade-off between minimizing $J(\boldsymbol{\theta})$ and $G(\boldsymbol{\theta})$, while allowing a maximum relaxation $c^*$ of the reference $G(\boldsymbol{\theta}_G^*)$. The weight parameter $w$ governs this trade-off: for sufficiently small $w$, the objective in \eqref{eq_opti_problem} emphasizes minimizing $J(\boldsymbol{\theta})$, whereas for sufficiently large $w$, the emphasis shifts toward minimizing $G(\boldsymbol{\theta})$. Section~\ref{sec_ills_ex} provides simulation results that empirically validate this behavior. The detailed connection between \eqref{eq_opti_problem} and policy adaptation in RL is discussed in the following subsection. 
\begin{remark}
    In the context of policy adaptation in RL, the formulation in \eqref{eq_opti_problem} differs fundamentally from conventional unconstrained multi-objective optimization approaches, which are typically formulated as \begin{equation}\label{eq_mutirl}
        \min_{\boldsymbol{\theta}\in\mathbb{R}^p} J(\boldsymbol{\theta}) + w\parallel G(\boldsymbol{\theta}_G^*) - G(\boldsymbol{\theta})\parallel^2.
    \end{equation}
    This objective incorporates a penalty term to discourage deviation from the reference solution $G(\boldsymbol{\theta}_G^*)$, where $\boldsymbol{\theta}$ is often initialized via warm-start parameters $\boldsymbol{\theta}=\boldsymbol{\theta}_G^*$. In contrast, the proposed problem in \eqref{eq_opti_problem} introduces a relaxation variable $c$ to treat the deviation from $G(\boldsymbol{\theta}_G^*)$ as a constraint, while minimizing a quadratic penalty on the maximum deviation, i.e., $c^2$. This formulation explicitly characterizes the trade-off between minimizing $J(\boldsymbol{\theta})$ and $G(\boldsymbol{\theta})$ through the tolerance variable $c$, which can be regulated using standard constrained optimization methods. As demonstrated in Section~\ref{sec:sim_sim}, for the same penalty weight $w$, the proposed formulation \eqref{eq_opti_problem} yields smaller deviations from $G(\boldsymbol{\theta}_G^*)$ compared to \eqref{eq_mutirl}. 
\end{remark}

\subsection{Problem Interpretation as Policy Adaptation in RL}
We consider a standard RL setup, where an agent interacts with an environment at discrete-time steps $t=0,1,2,\cdots$. At any time $t$, given a system state $\boldsymbol{x}(t)\in\mathcal{X}\subset\mathbb{R}^n$, the agent selects the control input $\boldsymbol{u}(t)\in\mathcal{U}\subset\mathbb{R}^m$ according to a parameterized policy/actor: \begin{equation}\label{eq_newpolicy}
    \boldsymbol{u}(t) = \boldsymbol{\mu}(\boldsymbol{x}(t), \boldsymbol{\theta}^{\mu}),
\end{equation} where $\boldsymbol{\mu}(\cdot,\boldsymbol{\theta}^{\mu}): \mathcal{X}\rightarrow\mathcal{U}$ represents a function with a known structure parameterized by $\boldsymbol{\theta}^{\mu}\in\mathbb{R}^p$. Upon executing the $\boldsymbol{u}(t)$, the agent moves to a new state $\boldsymbol{x}(t+1)\sim\mathcal{P}(\cdot|\boldsymbol{x}(t), \boldsymbol{u}(t))$, where $\mathcal{P}:\mathcal{X}\times\mathcal{U}\rightarrow\mathcal{X}$ denotes an unknown Markov transition kernel that describes the evolution of the system states. In the rest of this paper, we denote $\boldsymbol{x}_t$ and $\boldsymbol{u}_t$ as constant state and input vectors, respectively, to distinguish them from the state and input variables.
\subsubsection{Policy Adaptation Problem in RL}
Objectives of actor-critic algorithms \cite{sutton1988learning} usually involve finding a policy that minimizes the discounted sum of stage costs, i.e., \begin{equation}\label{eq_R_ori}
\min_{\boldsymbol{\theta}^\mu\in\mathbb{R}^p} G(\boldsymbol{\theta}^\mu) = \underset{\boldsymbol{x}(t)\sim\mathcal{P}_0, \boldsymbol{u}(s)\sim\boldsymbol{\mu}}{\mathbb{E}}[\sum_{s=t}^T \hat\gamma^{s-t} \hat\phi(\boldsymbol{x}(s),\boldsymbol{u}(s))],
\end{equation}
where $\mathcal{P}_0$ denotes the distribution of the initial state $\boldsymbol{x}(t)$, $\boldsymbol{u}(s)$ is the control input generated from the policy in \eqref{eq_newpolicy}, $0<\hat\gamma\leq 1$ is a discount factor, and $\hat\phi(\boldsymbol{x}(s), \boldsymbol{u}(s)): \mathcal{X}\times\mathcal{U}\rightarrow\mathbb{R}$ is the stage cost function.

Since computing $G(\boldsymbol{\theta}^\mu)$ in \eqref{eq_R_ori} requires knowledge of the unknown transition kernel $\mathcal{P}$, actor–critic methods approximate the optimal policy that minimizes $G(\boldsymbol{\theta}^\mu)$ in a model-free manner. This is achieved by defining the action-value function $\hat{Q}_{\mu}(\boldsymbol{x}_t,\boldsymbol{u}_t)$ for $G(\boldsymbol{\theta}^\mu)$, which quantifies the expected return starting from a fixed state–action pair $(\boldsymbol{x}_t, \boldsymbol{u}_t)$ and subsequently following policy $\boldsymbol{\mu}$ in \eqref{eq_newpolicy}. It is defined as
\begin{equation}\label{eq_action_value_ori}
\hat{Q}_{\mu}(\boldsymbol{x}_t,\boldsymbol{u}_t) = \underset{\boldsymbol{x}_{s> t}\sim\mathcal{P}, \boldsymbol{u}_{s>t}\sim\boldsymbol{\mu}}{\mathbb{E}}[\sum_{s=t}^T \hat\gamma^{s-t} \hat\phi(\boldsymbol{x}(s),\boldsymbol{u}(s)) | \boldsymbol{x}_t, \boldsymbol{u}_t]. \nonumber
\end{equation}
Then, an optimal critic function \cite{watkins1992q}, $\hat{Q}(\boldsymbol{x}_t,\boldsymbol{u}_t, \boldsymbol{\theta}^{\hat Q *})$, is used to approximate the action-value function, where $\hat{Q}(\cdot,\cdot, \boldsymbol{\theta}^{\hat Q *}):\mathcal{X}\times\mathcal{U}\rightarrow\mathbb{R}$ is defined with known structure and constant parameters $\boldsymbol{\theta}^{\hat Q *}\in\mathbb{R}^{\hat q}$, such that \begin{equation}\label{eq_optimal_pre_critic}
\hat{Q}(\boldsymbol{x}_t,\boldsymbol{u}_t, \boldsymbol{\theta}^{\hat Q *}) = \hat{Q}_{\mu}(\boldsymbol{x}_t,\boldsymbol{u}_t).
\end{equation}
Finally, the optimal policy that minimizes \eqref{eq_R_ori} is indirectly obtained by minimizing the critic evaluation: \begin{equation}\label{eq_optimal_pre_actor}
\boldsymbol{\theta}^{\hat\mu *} = \arg\min_{\boldsymbol{\theta}^\mu\in\mathbb{R}^p}\mathbb{E}_{\boldsymbol{x}_t\sim\rho_{\mu}}[\hat{Q}(\boldsymbol{x}_t, \boldsymbol{\mu}(\boldsymbol{x}_t,\boldsymbol{\theta}^\mu), \boldsymbol{\theta}^{\hat Q*})],
\end{equation}
where $\rho_{\mu}$ denotes the known stationary discounted state visitation distribution induced by policy $\boldsymbol{\mu}$.

Given that the optimal policy for minimizing the original objective in \eqref{eq_R_ori} has been obtained using actor–critic methods, i.e., the optimal actor $\boldsymbol{\mu}(\cdot,\boldsymbol{\theta}^{\hat\mu *})$ in \eqref{eq_optimal_pre_actor} is available. The policy adaptation problem in RL seeks to further tune $\boldsymbol{\theta}^{\hat\mu *}$ to obtain a trade-off solution between minimizing the original objective in \eqref{eq_R_ori} and a secondary optimization problem defined as:
\begin{equation}\label{eq_R_add}
\min_{\boldsymbol{\theta}^{\mu}\in\mathbb{R}^p} J(\boldsymbol{\theta}^\mu) = \mathbb{E}_{\boldsymbol{x}(t)\sim\mathcal{P}_0, \boldsymbol{u}(s)\sim\boldsymbol{\mu}}[\sum_{s=t}^T \gamma^{s-t} \phi(\boldsymbol{x}(s),\boldsymbol{u}(s))],
\end{equation}
where $0<\gamma\leq 1$ is the discount factor, $\phi(\boldsymbol{x}(s), \boldsymbol{u}(s)): \mathcal{X}\times\mathcal{U}\rightarrow\mathbb{R}$ denotes the stage cost function, and control input $\boldsymbol{u}(s)$ is generated by \eqref{eq_newpolicy} initialized at $\boldsymbol{\theta}^\mu \coloneqq\boldsymbol{\theta}^{\hat\mu *}$.

\subsubsection{Connection Between the Proposed Constrained Optimization and Policy Adaptation in RL}
The policy adaptation problem in RL can be cast as the constrained optimization in \eqref{eq_opti_problem} by setting $\boldsymbol{\theta}_G^*\coloneqq\boldsymbol{\theta}^{\hat\mu *}$ as the parameters of the predesigned optimal policy, $\boldsymbol{\theta}\coloneqq\boldsymbol{\theta}^\mu$ as the adaptable policy parameters, and $G(\boldsymbol{\theta})$ and $J(\boldsymbol{\theta})$ as the original objective in \eqref{eq_R_ori} and the additional objective in \eqref{eq_R_add}, respectively. Specifically, this yields the optimization problem \begin{equation}\label{eq_opti_problem_rl}
    \begin{aligned}
\min_{(\boldsymbol{\theta}^\mu,c)\in\mathbb{R}^{p+1}} &J(\boldsymbol{\theta}^\mu) + wc^2\\
\text{subject to} \quad &G(\boldsymbol{\theta}^{\hat{\mu} *}) - G(\boldsymbol{\theta}^\mu) + c\geq 0, \ c \geq 0, \nonumber
    \end{aligned}
\end{equation}
where $\boldsymbol{\theta}^{\hat{\mu} *}$ denotes the pretrained optimal policy parameters obtained from \eqref{eq_optimal_pre_actor}, $G(\boldsymbol{\theta}^\mu)$ is the original objective defined in \eqref{eq_R_ori}, and $J(\boldsymbol{\theta}^\mu)$ as the secondary objective defined in \eqref{eq_R_add}.

\section{Main Results}\label{sec:main}
In this section, we first highlight the key challenges and principles underlying the solution of \eqref{eq_opti_problem}. Building on these insights, we introduce a closed-loop dynamical system to address \eqref{eq_opti_problem} and provide a theoretical analysis establishing constraint satisfaction throughout the optimization process.

\subsection{Challenges and Key Ideas}
Recent studies \cite{allibhoy2023, hauswirth2024optimization, raghunathan2025constrained} have explored how solutions to constrained optimization problems can be interpreted as closed-loop control laws acting on the dynamics of the decision variables. This control-oriented perspective models optimization algorithms as closed-loop systems, enabling the application of stability, invariance, and robustness theory to algorithm design.

Motivated by these developments, we consider the following continuous-time parameter dynamics to solve \eqref{eq_opti_problem}:
\begin{align}
    \boldsymbol{\dot \theta} &= -\nabla_{\boldsymbol{\theta}}J(\boldsymbol{\theta}) + \boldsymbol{a}(\boldsymbol{\theta},c^*), \label{eq_dyn_theta} \\
    \dot{c}^* &= \psi(\boldsymbol{\theta}), \label{eq_dyn_c}
\end{align}
where $\boldsymbol{a}(\boldsymbol{\theta},c^*):\mathbb{R}^p\times\mathbb{R}\rightarrow\mathbb{R}^p$ denotes a closed-loop controller introduced to enforce the constraint in \eqref{eq_presol_cons}, and $\psi(\boldsymbol{\theta}):\mathbb{R}^p\rightarrow\mathbb{R}$ characterizes the dynamics of the minimal relaxation constant $c^*$, to be determined. We assume that the gradient $\nabla_{\boldsymbol{\theta}}J(\boldsymbol{\theta})$ is locally Lipschitz continuous. Since $\boldsymbol{a}(\boldsymbol{\theta},c^*)$ depends on both $c^*$ and $\boldsymbol{\theta}$, and $c^*$ itself is determined by $\boldsymbol{\theta}$, we adopt the shorthand notation $\boldsymbol{a}(\boldsymbol{\theta})\coloneqq \boldsymbol{a}(\boldsymbol{\theta},c^*)$ in the rest of this paper. For brevity, given any $c^*\geq 0$, we define the constraint-admissible set associated with  \eqref{eq_presol_cons} as \begin{equation}\label{eq_safeset_theta}
    \mathcal{C}_{\theta, c^*}=\{\boldsymbol{\theta}\in\mathbb{R}^p: h(\boldsymbol{\theta}) + c^*\geq 0\}.
\end{equation}
Three main challenges arise in the design of $\boldsymbol{a}(\boldsymbol{\theta})$ in \eqref{eq_dyn_theta}.

\noindent 1) When $h(\boldsymbol{\theta})$ is highly nonlinear, deriving a closed-form expression for $\boldsymbol{a}(\boldsymbol{\theta})$ that enforces the constraint \eqref{eq_presol_cons} is difficult.

\noindent 2) Since the constraint in \eqref{eq_presol_cons} is defined directly in terms of the parameter $\boldsymbol{\theta}$, it is necessary to construct a feasible set for the parameter dynamics $\boldsymbol{\dot \theta}$ in \eqref{eq_dyn_theta}, denoted as $\mathcal{C}_{\dot\theta, c^*}$, such that if $\boldsymbol{\theta}$ is initialized at $\boldsymbol{\theta}_G^*$ and  $\boldsymbol{\dot\theta}\in\mathcal{C}_{\dot\theta, c^*}$, then $\boldsymbol{\theta}\in\mathcal{C}_{\theta, c^*}$.

\noindent 3) Efficiently determining the dynamics corresponding to the minimal relaxation constant \eqref{eq_dyn_c} is nontrivial.

To address these challenges, we construct the feasible set $\mathcal{C}_{\dot\theta, c^*}$ for the dynamics in \eqref{eq_dyn_theta} based on the set-invariance property of continuous-time control barrier functions (CBFs), which enables the nonlinear constraint \eqref{eq_presol_cons} to be reformulated as a linear constraint without any relaxation. Based on this formulation, we design a quadratic programming (QP) that simultaneously determines $\boldsymbol{a}(\boldsymbol{\theta})$ and $c^*$.

\textbf{Feasible set construction for parameter update direction.} Given the parameter dynamics in \eqref{eq_dyn_theta}, we have the following definitions related to CBFs along with their fundamental property:
\begin{definition}[CBFs for Continuous-time Dynamical Systems \cite{ames2019control}]\label{def1}
  A continuously differentiable function $B(\boldsymbol{\theta}):\mathbb{R}^p\rightarrow \mathbb{R}, B\geq 0$ is a CBF if there exists an extended class $\mathcal{K}$ function $\kappa$ such that $\forall \boldsymbol{\theta}\in\mathbb{R}^p$, it meets the following condition:
\begin{equation}\label{eq_dis_cbf}
    \sup_{\boldsymbol{a}\in\mathbb{R}^p}\{L_f B(\boldsymbol{\theta}) + L_g B(\boldsymbol{\theta})\boldsymbol{a}\}\geq - \kappa(B(\boldsymbol{\theta})),
\end{equation}
\end{definition}
where $L_f B(\boldsymbol{\theta}) = -\frac{\partial B(\boldsymbol{\theta})}{\partial \boldsymbol{\theta}}\nabla_{\boldsymbol{\theta}}J(\boldsymbol{\theta})\in\mathbb{R}$ and $L_g B(\boldsymbol{\theta}) = \frac{\partial B(\boldsymbol{\theta})}{\partial \boldsymbol{\theta}}\in\mathbb{R}^{1\times p}$ are Lie derivatives.
Accordingly, we define the superlevel set of $h$ as the safe set $\mathcal{C}$, given by: \begin{equation}\label{eq_safe_set}
    \mathcal{C} = \{\boldsymbol{\theta}\in\mathbb{R}^p: B(\boldsymbol{\theta})\geq 0\}.
\end{equation} \begin{definition}[Constraint Satisfaction with CBFs \cite{ames2019control}]\label{def2}
Suppose the dynamical system is defined as in \eqref{eq_dyn_theta}, with a CBF $B(\boldsymbol{\theta})$ specified in Definition \ref{def1} and the corresponding safe set $\mathcal{C}$ given in \eqref{eq_safe_set}. Any Lipschitz continuous controller that satisfies \eqref{eq_dis_cbf} ensures that the safe set $\mathcal{C}$ remains invariant, thus guaranteeing constraint satisfaction of $B(\boldsymbol{\theta})\geq 0$.
\end{definition} 

According to Definitions~\ref{def1}-\ref{def2}, we define the feasible set  $\mathcal{C}_{\boldsymbol{\dot\theta}, c^*}$ for $\boldsymbol{\dot\theta}$ within the CBFs framework by treating $h(\boldsymbol{\theta}) + c^*$ as a candidate CBF. The validity of this candidate will be formally established in Section~\ref{subsec:analysis}. By Definition~\ref{def2}, any $\boldsymbol{a}(\boldsymbol{\theta})$ that satisfies the CBF inequality guarantees forward invariance of the set $\mathcal{C}_{\theta, c^*}$ in \eqref{eq_safeset_theta}. Consequently, for any $c^*\geq 0$, we introduce the following CBF inequality constraint:
\begin{equation}\label{eq_cbf_constraint}
   \underbrace{L_f  + L_g \boldsymbol{a} + \kappa(h(\boldsymbol{\theta})+c^*)}_{\hat{g}(\boldsymbol{a},c^*)}\geq 0,
\end{equation}
where \begin{equation}\label{eq_Lh}
    \begin{aligned}
        L_f &= -\nabla_{\boldsymbol{\theta}}h(\boldsymbol{\theta})'\nabla_{\boldsymbol{\theta}}J(\boldsymbol{\theta})= \nabla_{\boldsymbol{\theta}}G(\boldsymbol{\theta})' \nabla_{\boldsymbol{\theta}}J(\boldsymbol{\theta}),\\
    L_g &=\nabla_{\boldsymbol{\theta}}h(\boldsymbol{\theta})'=-\nabla_{\boldsymbol{\theta}}G(\boldsymbol{\theta})'
    \end{aligned}
\end{equation}
are Lie derivatives of the function $h(\boldsymbol{\theta})+c^*$ under the parameter dynamics given in \eqref{eq_dyn_theta}. Furthermore, by applying \eqref{eq_dyn_theta}, $\hat{g}(\boldsymbol{a},c^*)$ in \eqref{eq_cbf_constraint} can be equivalently written as: \begin{equation}
    \hat{g}(\boldsymbol{a},c^*) =\nabla_{\boldsymbol{\theta}}h(\boldsymbol{\theta})'\boldsymbol{\dot \theta} + \kappa(h(\boldsymbol{\theta})+c^*), \nonumber
\end{equation} which leads to the definition of the following feasible set of $\boldsymbol{\dot\theta}$: \begin{equation}\label{eq_safeset_dottheta}
\mathcal{C}_{\boldsymbol{\dot\theta}, c^*} = \{\boldsymbol{\dot \theta}\in\mathbb{R}^p: \nabla_{\boldsymbol{\theta}}h(\boldsymbol{\theta})'\boldsymbol{\dot \theta} + \kappa(h(\boldsymbol{\theta})+c^*) \geq 0\}. 
\end{equation} \begin{remark}\label{rm3}
    As shown in \eqref{eq_safeset_dottheta}, finding the control input $\boldsymbol{a}$ in \eqref{eq_cbf_constraint} is equivalent to determining the parameters update direction $\boldsymbol{\dot\theta}\in\mathcal{C}_{\boldsymbol{\dot\theta}, c^*}$. Furthermore, according to the properties of CBFs stated in Definition~\ref{def2}, if the parameters $\boldsymbol{\theta}$ is initialized within the set  $\mathcal{C}_{\theta, c^*}$, then any update direction $\boldsymbol{\dot\theta}\in\mathcal{C}_{\boldsymbol{\dot\theta}, c^*}$ guarantees that the corresponding parameter trajectories remains within $\boldsymbol{\theta}\in\mathcal{C}_{\theta, c^*}$ throughout the evolution. 
\end{remark}  

\textbf{Construct the QP problem.} Consider the parameter dynamics in \eqref{eq_dyn_theta} and the feasible set specified in \eqref{eq_safeset_dottheta}. We construct the following QP problem to derive $\boldsymbol{a}(\boldsymbol{\theta})$ and $c^*$:
\begin{equation}\label{eq_cbfqp}
\begin{aligned}
(\boldsymbol{a}(\boldsymbol{\theta}), c^*) = \arg\min_{(\boldsymbol{a},c)\in\mathbb{R}^{p+1}} &\frac{1}{2}\parallel \boldsymbol{a}\parallel^2  + \frac{w}{2} c^2\\
\text{subject to}\quad &\hat{g}(\boldsymbol{a},c)\geq 0, \  c\geq 0.
\end{aligned}
\end{equation}
\begin{remark}
For any $\boldsymbol{\theta}\in\mathbb{R}^p$, \eqref{eq_cbfqp} is formulated to derive the closed-loop controller $\boldsymbol{a}(\boldsymbol{\theta})$ and the minimal relaxation constant $c^*$ that enforce the CBF inequality in \eqref{eq_cbf_constraint}, thereby guaranteeing satisfaction of constraint \eqref{eq_presol_cons}, while simultaneously minimizing both quantities.
\end{remark}

\subsection{The Proposed Closed-Loop System}
Consider the parameter dynamics in \eqref{eq_dyn_theta}, where the parameter is initialized as $\boldsymbol{\theta} = \boldsymbol{\theta}_G^*$.  We introduce the proposed closed-loop controller $\boldsymbol{a}(\boldsymbol{\theta})$ for \eqref{eq_dyn_theta} to solve \eqref{eq_opti_problem}, which corresponds to the closed-form solution of \eqref{eq_cbfqp}, as formalized in the following lemma. In the remainder of this paper, we refer to dynamics \eqref{eq_dyn_theta} with proposed $\boldsymbol{a}(\boldsymbol{\theta})$ as the \textit{control-barrier-function-based policy adaptation} (CBF-PA) algorithm.
\begin{lemma}\label{lemma1}
For any $\boldsymbol{\theta}\in\mathbb{R}^p$, if $\kappa(h(\boldsymbol{\theta}) + c) = \gamma_{\mathrm{h}}(h(\boldsymbol{\theta})+c)$ with $\gamma_{\mathrm{h}}>0$ a given constant, then the closed-loop controller $\boldsymbol{a}(\boldsymbol{\theta})$ and minimal relaxation constant $c^*$ that solve \eqref{eq_cbfqp} are:
\begin{equation}\label{eq_alg_qp_sol}
   (\boldsymbol{a}(\boldsymbol{\theta}), c^*) = \begin{cases}
   (\mathbf{0}_p, 0),
      &\text{if}\ L_a\geq 0, \\ (\frac{-L_a L_g'}{L_g L_g' + \gamma_{\mathrm{h}}^2/w},L_b), &\text{if}\ L_a< 0,   L_b\geq 0,\\ (\frac{-L_a L_g'}{L_g L_g'}, 0),
      &\text{if}\ L_a < 0, L_b< 0,
    \end{cases} 
\end{equation} 
where the auxiliary terms are given by
\begin{equation}
    \begin{aligned}
        L_a &\coloneqq L_f  + \gamma_{\mathrm{h}}(G(\boldsymbol{\theta}_G^*) - G(\boldsymbol{\theta})),\\
        L_b &\coloneqq -\gamma_{\mathrm{h}}L_a/(w L_gL_g' +\gamma_{\mathrm{h}}^2),  \nonumber
    \end{aligned}
\end{equation}
and $L_f$ and $L_g$ are defined in \eqref{eq_Lh}.
\end{lemma} The proof of Lemma~\ref{lemma1} is provided in the Appendix.
\begin{remark}
    For the special case where $w=0$ and $c$ is a given constant. Under the same assumptions and notations as in Lemma~\ref{lemma1}, the following closed-form solution holds: \begin{equation}
   \boldsymbol{a}(\boldsymbol{\theta}) = \begin{cases}
   \mathbf{0}_p,
      &\text{if}\ L_a + \gamma_{\mathrm{h}} c \geq 0, \\ \frac{-(L_a + \gamma_{\mathrm{h}} c) L_g'}{L_g L_g'},
      &\text{else}.
    \end{cases} \nonumber
\end{equation} The proof proceeds analogously to the  proof of Lemma~\ref{lemma1} and is thus omitted for brevity.
\end{remark}

\subsection{Analysis}\label{subsec:analysis}
This subsection presents a theoretical analysis of the proposed closed-loop controller in Lemma~\ref{lemma1} for solving the constrained optimization problem in \eqref{eq_opti_problem}.
Since this paper utilizes the properties of CBFs to ensure \eqref{eq_presol_cons} is satisfied, it is essential to verify that $\forall c\geq 0$, the function $h(\boldsymbol{\theta})+c$ in \eqref{eq_presol_cons} is a valid CBF for $\mathcal{C}_{\theta, c}$. To this end, we establish the following: 
\begin{lemma}\label{lemma_cbf_valid}
    Given the parameter dynamics in \eqref{eq_dyn_theta}, for any $c\geq 0$, $h(\boldsymbol{\theta})+c$ in \eqref{eq_presol_cons} is a valid CBF for $\mathcal{C}_{\theta, c}$ in \eqref{eq_safeset_theta}.
\end{lemma} The proof of Lemma~\ref{lemma_cbf_valid} is given in the Appendix.
The following theorem establishes the constraint satisfaction associated with the optimal closed-loop controller in \eqref{eq_alg_qp_sol}:
\begin{theorem}\label{thm_a_star}
    Consider the parameter dynamics \eqref{eq_dyn_theta} with $\boldsymbol{a}(\boldsymbol{\theta})$ and $c^*$ defined in \eqref{eq_alg_qp_sol}. Suppose $\boldsymbol{\theta}$ is initialized as $\boldsymbol{\theta} = \boldsymbol{\theta}_G^*$, $\nabla_{\boldsymbol{\theta}}J(\boldsymbol{\theta})$ in \eqref{eq_dyn_theta} is locally Lipschitz continuous and bounded, and the class-$\mathcal{K}$ function $\kappa(h(\boldsymbol{\theta})+c^*)$ satisfies $\kappa(h(\boldsymbol{\theta})+c^*) = \gamma_{\mathrm{h}}(h(\boldsymbol{\theta})+c^*)$ with $\gamma_{\mathrm{h}} > 0$. Then, if $\boldsymbol{\theta}$ evolves according to \eqref{eq_dyn_theta}, it follows that $\boldsymbol{\theta}\in\mathcal{C}_{\theta, c^*}$ as defined in \eqref{eq_safeset_theta}. 
\end{theorem} The proof of Theorem~\ref{thm_a_star} is provided in the Appendix.

\subsection{Applying the Proposed CBF-PA to Policy Adaptation in RL} We now propose an algorithm that integrates the CBF-PA mechanism to address the policy adaptation problem in RL. The approach extends the deep deterministic policy gradient (DDPG) method \cite{lillicrap2015continuous} by employing CBF-PA as the policy update rule. In contrast to standard DDPG, the proposed CBF-PA algorithm refines the learned policy to minimize an additional task objective while preserving near-optimal performance on the original task. This refinement is achieved by augmenting the policy gradient with the analytically derived closed-loop controller in Lemma~\ref{lemma1}.

The design builds upon the DDPG framework, where the optimal critic $Q(\boldsymbol{x}_t,\boldsymbol{u}_t, \boldsymbol{\theta}^{Q*})$ is used to approximate the action-value function of $J(\boldsymbol{\theta}^\mu)$ in \eqref{eq_R_add}. The optimal policy parameters  $\boldsymbol{\theta}^{\mu *}$ is then obtained by minimizing the critic feedback, i.e., $\boldsymbol{\theta}^{\mu *} = \arg\min_{\boldsymbol{\theta}^{\mu}\in\mathbb{R}^p}\mathbb{E}_{\boldsymbol{x}_t\sim\rho_{\mu}}[Q(\boldsymbol{x}_t, \boldsymbol{\mu}(\boldsymbol{x}_t,\boldsymbol{\theta}^{\mu}), \boldsymbol{\theta}^{Q*})]$, where $\rho_{\mu}$ denotes the discounted state visitation distribution under policy $\boldsymbol{\mu}$.

To initialize the proposed algorithm, it is assumed that a pretrained optimal critic $\hat{Q}(\cdot, \cdot, \boldsymbol{\theta}^{\hat Q *})$, as defined in \eqref{eq_optimal_pre_critic}, and an optimal policy $\boldsymbol{\mu}(\cdot, \boldsymbol{\theta}^{\hat\mu *})$, as specified in \eqref{eq_optimal_pre_actor}, are already available. The learning process begins by constructing a new critic function $Q(\cdot,\cdot, \boldsymbol{\theta}^Q):\mathcal{X}\times\mathcal{U}\rightarrow\mathbb{R}$, and initializing a policy $\boldsymbol{\mu}(\cdot, \boldsymbol{\theta}^\mu)$, where $\boldsymbol{\mu}$ and $\boldsymbol{\mu}$ share the same functional structure, with the initial policy parameters set as $\boldsymbol{\theta}^\mu = \boldsymbol{\theta}^{\hat\mu*}$. 

Building upon the actor-critic framework, we propose an iterative method for estimating the optimal parameters $\boldsymbol{\theta}^{Q *}$ and $\boldsymbol{\theta}^{\mu *}$. Recall $k=0,1,2,\cdots$ denote the iteration index, and let $\boldsymbol{\theta}_k^Q$ and $\boldsymbol{\theta}_k^\mu$ represent the estimates of $\boldsymbol{\theta}^{Q *}$ and $\boldsymbol{\theta}^{\mu *}$ at the $k$-th iteration, respectively. The constant step sizes for updating the critic and actor parameters are denoted by $\alpha_Q$ and $\alpha_\mu = \epsilon\alpha_Q$, respectively, where $0<\epsilon<1$ is a constant used to ensure that the critic update proceeds at a faster rate than the actor update. The following iterative update rules are employed to refine these estimates:
\subsubsection{Critic Update}
 To find the optimal critic that approximates the action-value function of the objective function in \eqref{eq_R_add}, the critic parameters  $\boldsymbol{\theta}_k^Q$ is updated as follows \cite{watkins1992q}:
\begin{equation}\label{eq_alL_britic}
  \boldsymbol{\theta}_{k+1}^Q = \boldsymbol{\theta}_k^Q - \alpha_Q \nabla_{\boldsymbol{\theta}^Q}\bL_Q(\boldsymbol{\theta}_k^Q, \boldsymbol{\theta}_k^\mu), \quad \boldsymbol{\theta}_0^Q, \boldsymbol{\theta}_0^\mu\ \text{given},
\end{equation} where $\bL_Q(\boldsymbol{\theta}_k^Q, \boldsymbol{\theta}_k^\mu)$ denotes the temporal difference learning loss function \cite{sutton1988learning} defined by \begin{equation}
\begin{aligned}
    \bL_Q(\boldsymbol{\theta}_k^Q, \boldsymbol{\theta}_k^\mu) = \mathbb{E}_{\boldsymbol{x}_t\sim\rho_{\mu}, \boldsymbol{u}_t\sim\boldsymbol{\mu}}[(Q(\boldsymbol{x}_t, \boldsymbol{u}_t, \boldsymbol{\theta}_k^Q) - \phi(\boldsymbol{x}_t, \boldsymbol{u}_t) \\- Q(\boldsymbol{x}_{t+1},  \boldsymbol{\mu}(\boldsymbol{x}_{t+1}, \boldsymbol{\theta}_k^\mu), \boldsymbol{\theta}_k^Q))^2]. \nonumber
\end{aligned}
\end{equation}

\subsubsection{Policy Adaptation using CBF-PA}
After updating $\boldsymbol{\theta}_k^Q$ via \eqref{eq_alL_britic}, the policy parameters $\boldsymbol{\theta}_k^\mu$ is subsequently updated using the proposed CBF-PA, given by:
\begin{equation}\label{eq_safe_actor_ac}
\begin{aligned}
    \boldsymbol{\theta}_{k+1}^\mu = \boldsymbol{\theta}_k^\mu -\alpha_\mu\Big(\mathbb{E}_{\boldsymbol{x}_t\sim\rho_{\mu}}[\nabla_{\boldsymbol{\theta}^\mu}Q(\boldsymbol{x}_t, \boldsymbol{\mu}(\boldsymbol{x}_t,\boldsymbol{\theta}_k^\mu), \boldsymbol{\theta}_{k+1}^Q)] \\- \boldsymbol{a}(\boldsymbol{\theta}_k^\mu)\Big), 
\end{aligned}
\end{equation}
where $\boldsymbol{a}(\boldsymbol{\theta}_k^\mu)$ is computed by \eqref{eq_alg_qp_sol} with the components:
\begin{equation}
    \begin{aligned}
        L_f  &= \mathbb{E}_{\boldsymbol{x}_t\sim\rho_{\mu}}[\nabla_{\boldsymbol{\theta}^\mu}\hat{Q}(\boldsymbol{x}_t,\boldsymbol{\mu}(\boldsymbol{x}_t, \boldsymbol{\theta}_k^\mu),\boldsymbol{\theta}^{\hat Q*})'\\ &\quad \cdot\nabla_{\boldsymbol{\theta}^\mu}Q(\boldsymbol{x}_t, \boldsymbol{\mu}(\boldsymbol{x}_t,\boldsymbol{\theta}_k^\mu), \boldsymbol{\theta}_{k+1}^Q)] \nonumber\\
        L_g &= -\mathbb{E}_{\boldsymbol{x}_t\sim\rho_{\mu}}[\nabla_{\boldsymbol{\theta}^\mu}\hat{Q}(\boldsymbol{x}_t,\boldsymbol{\mu}(\boldsymbol{x}_t, \boldsymbol{\theta}_k^\mu),\boldsymbol{\theta}^{\hat Q*})'],
        \\
        L_a &= L_f  + \gamma_{\mathrm{h}}\mathbb{E}_{\boldsymbol{x}_t\sim\rho_{\mu}}[\hat{Q}(\boldsymbol{x}_t, \boldsymbol{\mu}(\boldsymbol{x}_t, \boldsymbol{\theta}^{\hat\mu *}), \boldsymbol{\theta}^{\hat Q *}) \\&\quad - \hat{Q}(\boldsymbol{x}_t, \boldsymbol{\mu}(\boldsymbol{x}_t, \boldsymbol{\theta}^{\mu}), \boldsymbol{\theta}^{\hat Q *})]. \nonumber
    \end{aligned}
\end{equation}
The complete procedure is summarized in Algorithm~\ref{alg}. 
\begin{algorithm2e}
\DontPrintSemicolon
\caption{Control-Barrier-Function-Based Policy Adaptation (CBF-PA)}\label{alg}
\textbf{Initialize:} Pretrained optimal policy $\boldsymbol{\mu}(\cdot,\boldsymbol{\theta}^{\hat\mu*})$ and optimal critic $\hat{Q}(\cdot,
\cdot,\boldsymbol{\theta}^{\hat Q*})$; Empty data memory; initialize $Q(\cdot,\cdot, \boldsymbol{\theta}^Q)$, $\boldsymbol{\mu}(\cdot, \boldsymbol{\theta}^\mu)$ with $\boldsymbol{\theta}^\mu = \boldsymbol{\theta}^{\hat\mu*}$, and the iteration index $k=0$; copy $\bar{Q} \gets Q, \ \boldsymbol{\bar\mu} \gets \boldsymbol{\mu}$\; 
Set the number of episodes $E$, task
horizon $T$, update weight $\tau$, batch size $N$, $\gamma_{\mathrm{h}}$, and step sizes $\alpha_Q$ and $\alpha_\mu$\;
\For{$\emph{episode} = 0, 1, 2, \cdots, E$}
{
Initialize random process $W$ for action exploration and randomize initial state $\boldsymbol{x}(0)$\;
\For{$t = 0, 1, 2, \cdots, T$}
{
Execute control input $\boldsymbol{u}(t) = \boldsymbol{\mu}(\boldsymbol{x}(t), \boldsymbol{\theta}_k^\mu)$ $+ W(t)$, observe resulting $\boldsymbol{x}(t+1)$, and store data tuple $(\boldsymbol{x}_t, \boldsymbol{u}_t, \boldsymbol{x}_{t+1})$ in data memory\;
Sample $N$ tuples uniformly from data memory\;
Update critic and actor following \eqref{eq_alL_britic} and \eqref{eq_safe_actor_ac}, respectively, wherein the expectation terms are approximated using the averaged values computed over the $N$ sampled tuples\;
Update the copied networks $\bar{Q}$ and $\bar{\mu}$:
\begin{equation}
\begin{split}
\boldsymbol{\theta}^{\bar Q} &\leftarrow \tau \boldsymbol{\theta}^Q + (1-\tau)\boldsymbol{\theta}^{\bar Q}, \\ \boldsymbol{\theta}^{\bar\mu}  &\leftarrow \tau \boldsymbol{\theta}^\mu+ (1-\tau)\boldsymbol{\theta}^{\bar\mu}. \nonumber
\end{split}
\end{equation}
}
}
\textbf{Return: } $\boldsymbol{\theta}^{\bar Q}, \boldsymbol{\theta}^{\bar\mu}$\;
\end{algorithm2e}

\subsection{Discussion of Discrete-Time Implementation}
Since $\lim_{\alpha_\mu\rightarrow 0}\frac{\boldsymbol{\theta}_{k+1}^\mu- \boldsymbol{\theta}_k^\mu}{\alpha_\mu} = \boldsymbol{\dot\theta}^\mu,$ the policy update rule in \eqref{eq_safe_actor_ac} can be interpreted, as $\alpha_\mu\rightarrow 0$, as the Euler discretization of the continuous-time dynamics \[\boldsymbol{\dot\theta}^\mu = -\mathbb{E}_{\boldsymbol{x}_t\sim\rho_{\mu}}[\nabla_{\boldsymbol{\theta}^\mu}Q(\boldsymbol{x}_t, \boldsymbol{\mu}(\boldsymbol{x}_t,\boldsymbol{\theta}^\mu), \boldsymbol{\theta}^Q)] + \boldsymbol{a}(\boldsymbol{\theta}^\mu).\] This scheme introduces a discretization error of order $\mathcal{O}(\alpha_\mu)$ per gradient descent step \cite{miyagawa2022toward}, thereby causing a modeling discrepancy due to the nonzero $\alpha_\mu$. To ensure that the discrete-time implementation in \eqref{eq_safe_actor_ac} using $\boldsymbol{a}(\boldsymbol{\theta}_k^\mu)$ preserves the performance guarantees of the continuous-time formulation in \eqref{eq_dyn_theta} with $\boldsymbol{a}(\boldsymbol{\theta}^\mu)$ in \eqref{eq_cbfqp}, we adopt the following result:
\begin{lemma}[Inter-Sample Safety Guarantees, Theorem $3$ \cite{gurriet_applied_safety}]\label{lemma_continuous_discrete}
  Recall the policy parameter dynamics from \eqref{eq_dyn_theta} and the CBF $B(\boldsymbol{\theta})$ defined in Definition~\ref{def1}. Suppose that $-\nabla_{\boldsymbol{\theta}}J(\boldsymbol{\theta})$ and  $\boldsymbol{a}\in\mathcal{A}\subset\mathbb{R}^p$ are bounded. Let $\Delta$ be a constant such that \begin{equation*}
    \Delta \geq \frac{\alpha}{2}L_B\sup_{\boldsymbol{\theta} \in \mathbb{R}^p, \boldsymbol{a} \in \mathcal{A}}\lVert -\nabla_{\boldsymbol{\theta}}J(\boldsymbol{\theta}) + \boldsymbol{a} \rVert,
    \end{equation*} where $L_B$ denotes the Lipschitz constant of $B(\boldsymbol{\theta})$. Then, it holds that the closed-loop controller $\boldsymbol{a}$ satisfying the following tightened CBF inequality constraint of \eqref{eq_cbf_constraint}:
    \begin{equation*}
        L_f B(\boldsymbol{\theta}) + L_g B(\boldsymbol{\theta})\boldsymbol{a} + \kappa(B(\boldsymbol{\theta}) - \Delta) \geq 0
    \end{equation*} will guarantee the invariance of the $\mathcal{C}$ in \eqref{eq_safe_set}.
\end{lemma}
Lemma~\ref{lemma_continuous_discrete} says that if the discrepancy between the continuous-time dynamics and its Euler-discretized counterpart is bounded, then the control performance guaranteed by the CBF condition in the continuous-time setting can be preserved in the discretized system by employing a more conservative CBF inequality constraint.

\section{Numerical Simulations}\label{sec:sim_sim}
In this section, the proposed CBF-PA is first applied to a simple illustrative example, facilitating a direct comparison with standard gradient descent and the multi-objective optimization in \eqref{eq_mutirl}. This example also provides insight into the impact of key parameters, including the penalty weight $w$, step size $\alpha$, and CBF parameter $\gamma_{\mathrm{h}}$, on optimization performance. Subsequently, Algorithm~\ref{alg} is evaluated in two benchmark simulation environments, Cartpole and lunar lander, from OpenAI Gym \cite{brockman2016openai}, to assess its effectiveness and generalizability. For additional comparison, the method is evaluated against two representative transfer learning approaches, thereby highlighting its relative advantages.

\subsection{An Illustration Example}\label{sec_ills_ex}
Let $\boldsymbol{\theta} = [x, y]' \in\mathbb{R}^2$. We define the additional and original objective functions as $J(\boldsymbol{\theta}) = \sin(x) + (y-8)^2$ and $G(\boldsymbol{\theta}) = x^3 + y^3, \text{s.t.,}\ x\geq 0, y\geq 0$, respectively. The goal is to solve the optimization problem $\boldsymbol{\theta}^* = \arg\min_{\boldsymbol{\theta}\in\mathbb{R}^2}J(\boldsymbol{\theta})$, starting from a known initialization $\boldsymbol{\theta} = \boldsymbol{\theta}_G^*$, where \[\boldsymbol{\theta}_G^* = \matt{0 & 0}^\top =  \arg\min_{\boldsymbol{\theta}\in\mathbb{R}^2}G(\boldsymbol{\theta}),\ \text{s.t.}\ x\geq 0, y\geq 0.\] Note that for any $\boldsymbol{\theta} = [-x, x]'$, one obtains $G([-x, x]')=G(\boldsymbol{\theta}_G^*)=0$. This setup is designed to evaluate the algorithm’s ability to preserve the performance on the pre-optimized task $G$ while minimizing $J$. Accordingly, the performance metrics are defined as additional cost function $J(\boldsymbol{\theta}_k)$, original cost function $G(\boldsymbol{\theta}_k)$, and average value of $G(\boldsymbol{\theta}_k)$ over $K$ iterations: \[\bar G = \textstyle\frac{1}{K}\sum_{k=1}^{K}G(\boldsymbol{\theta}_k).\] 

\textbf{Benchmarks.} This experiment examines how different constant weight parameters $w$ influence the performance metric of the discrete-time implementation of the proposed method, defined as \begin{equation}\label{eq_safe_actor}
\begin{aligned}
\boldsymbol{\theta}_{k+1} = \boldsymbol{\theta}_k - \alpha(\nabla_\theta J(\boldsymbol{\theta_k}) - \boldsymbol{a}(\boldsymbol{\theta}_k)),
\end{aligned}
\end{equation} where $\boldsymbol{a}(\boldsymbol{\theta}_k)$ is given in Lemma~\ref{lemma1}. We compare \eqref{eq_safe_actor} with two baseline methods: standard gradient descent (GD) and a multi-objective gradient descent (MOGD). The GD method follows the classical update rule: \[\boldsymbol{\theta}_{k+1} = \boldsymbol{\theta}_k - \alpha\nabla_\theta J(\boldsymbol{\theta_k}),\] while the MOGD baseline incorporates a regularization term involving $G$, and is defined as: \[\boldsymbol{\theta}_{k+1} = \boldsymbol{\theta}_k - \alpha\nabla_\theta (J(\boldsymbol{\theta_k})+w(G(\boldsymbol{\theta_k}) - G(\boldsymbol{\theta}^*))^2).\] For the proposed method, we set $\gamma_{\mathrm{h}} = 10$, and use a fixed step size $\alpha = 0.001$ and initial parameter $\boldsymbol{\theta}_G^*$ for all methods. By varying $w$, we evaluate the relative optimization behavior and the trade-offs between minimizing the primary loss $J$ and maintaining performance with respect to $G$.

\begin{figure*}
\centering
\begin{subfigure}[b]{0.24\textwidth}
\centering
\includegraphics[width=\textwidth]{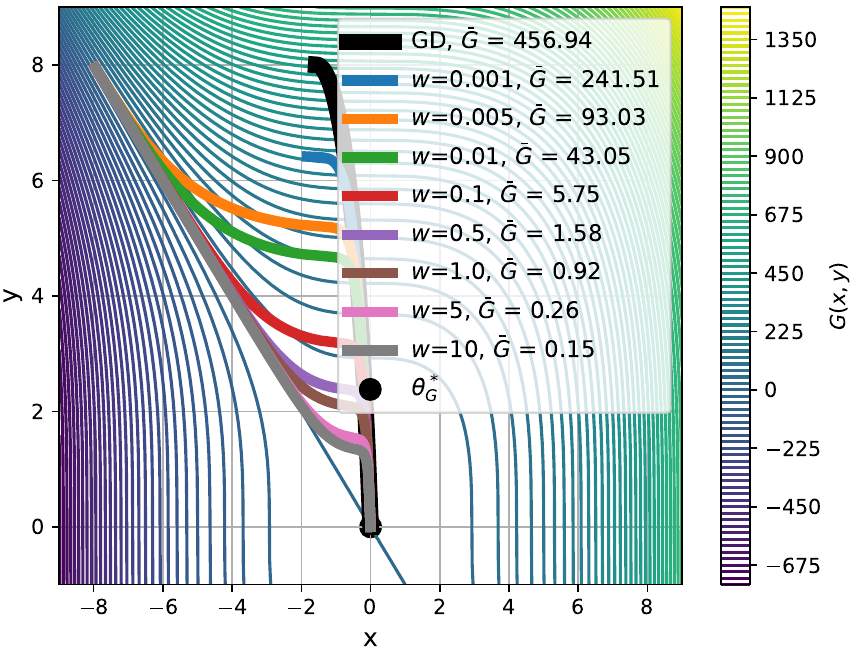}
\caption[]%
{{Parameter trajectories.}}    
\label{fig:com_g}
\end{subfigure}
\begin{subfigure}[b]{0.24\textwidth}
\centering
\includegraphics[width=\textwidth]{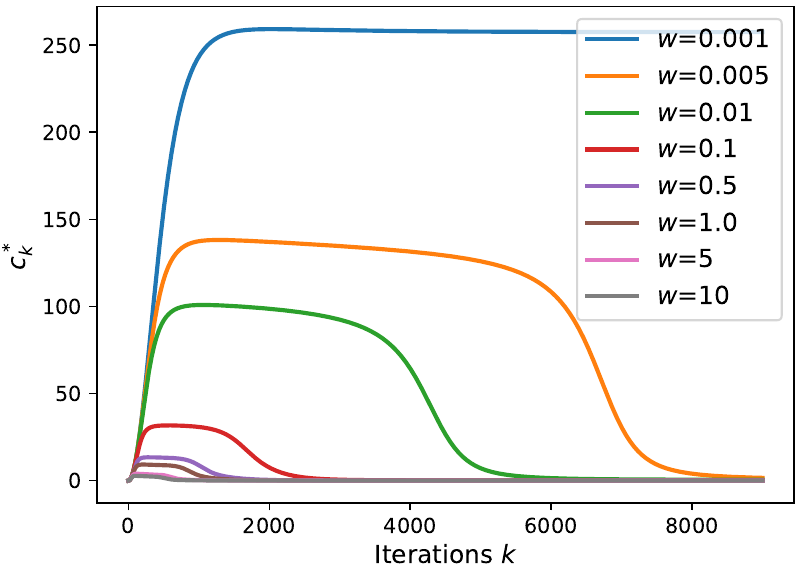}
\caption[]%
{{Evolution of $c_k^*$ of CBF-PA.}}
\label{fig:sim_c}
\end{subfigure}
\begin{subfigure}[b]{0.24\textwidth}
\centering
\includegraphics[width=\textwidth]{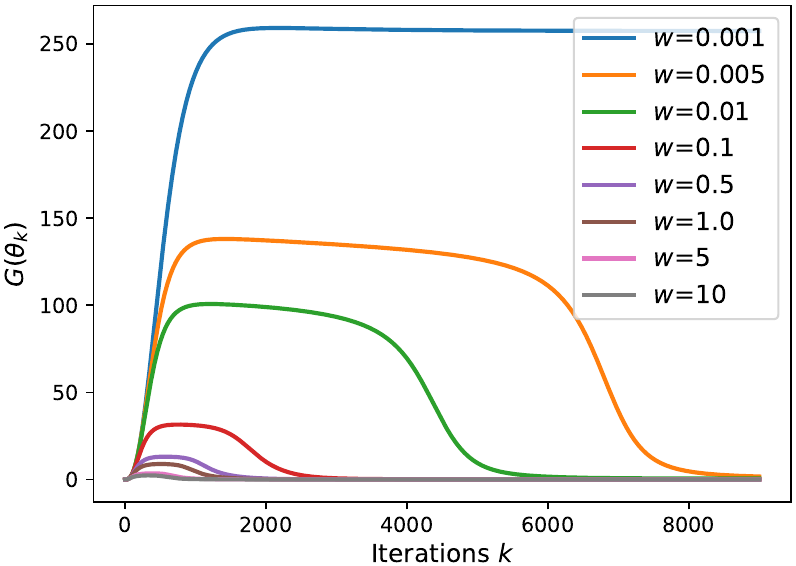}
\caption[]%
{{Original cost.}}    
\label{fig:com_G_cost_1}
\end{subfigure}
\begin{subfigure}[b]{0.24\textwidth}
\centering
\includegraphics[width=\textwidth]{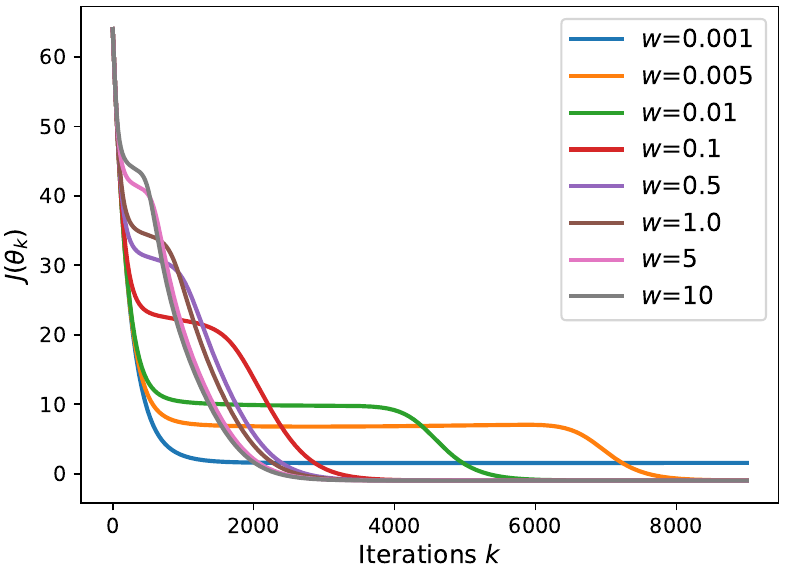}
\caption[]%
{{Additional cost.}}    
\label{fig:com_J_cost_1}
\end{subfigure}
\caption{Proposed CBF-PA under various values of $w$, shown over the contour plot of the function $G(x, y)$.} 
\label{fig:sim_com_cbfpa_w}
\end{figure*}

\begin{figure*}
\centering
\begin{subfigure}{0.30\textwidth}
\centering
\includegraphics[width=\textwidth]{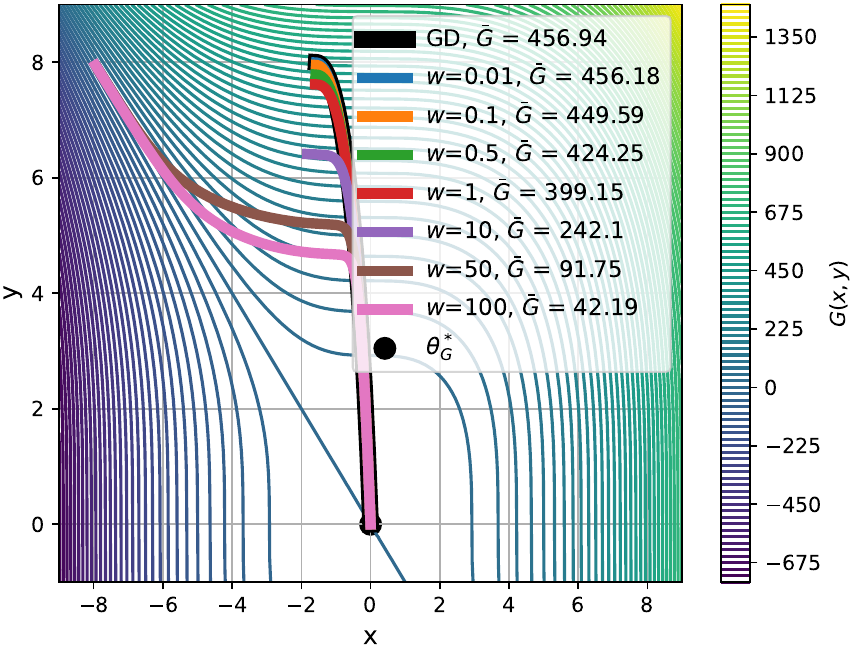}
\caption[]%
{{Parameter trajectories.}}
\label{fig:com_j}
\end{subfigure}
\begin{subfigure}{0.30\textwidth}
\centering
\includegraphics[width=\textwidth]{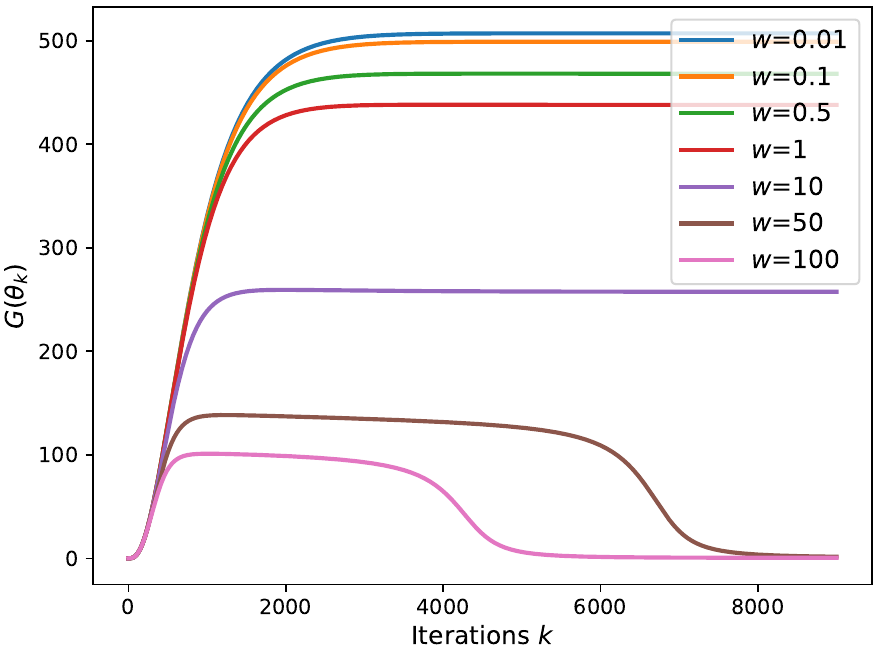}
\caption[]%
{{Original cost.}}
\label{fig:com_G_cost_2}
\end{subfigure}
\begin{subfigure}{0.30\textwidth}
\centering
\includegraphics[width=\textwidth]{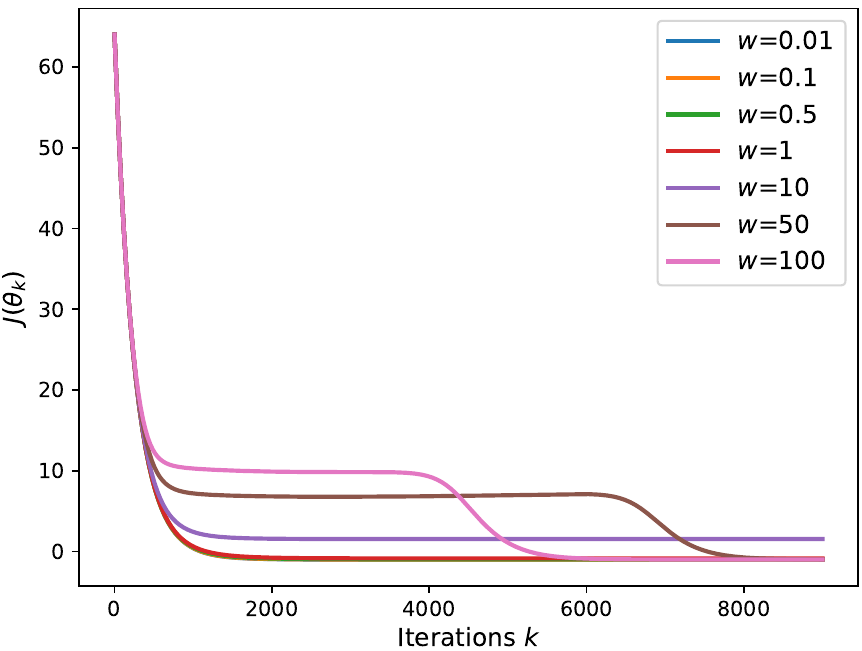}
\caption[]%
{{Additional cost.}}
\label{fig:com_J_cost_2}
\end{subfigure}
\caption{MOGD under various values of $w$, shown over the contour plot of the function $G(x, y)$.} 
\label{fig:sim_com_mogd}
\end{figure*}

\textbf{Results Analysis.}
As illustrated in Figs. \ref{fig:sim_com_cbfpa_w}-\ref{fig:sim_com_mogd}, the proposed method exhibits an optimization behavior similar to that of the MOGD algorithm under various weight parameters $w$. Specifically, decreasing $w$ drives the parameter trajectories of both methods toward the optimizer of the additional cost $J(\boldsymbol{\theta})$. Conversely, as $w$ increases, both methods yield parameter trajectories that approach the joint optimum of $J$ and $G$. Moreover, for larger values of $w$, the proposed CBF-PA demonstrates a faster convergence rate than MOGD.

\textbf{Key Parameters Investigation.} To evaluate the impact of key parameters in the proposed method \eqref{eq_safe_actor}, two sets of experiments are conducted. In the first set, the CBF parameter $\gamma_{\mathrm{h}}$ is varied while fixing the step size $\alpha = 0.001$ and penalty weight $w = 0.01$. In the second set, the step size $\alpha$ is varied with $\gamma_{\mathrm{h}} = 10$ and $w = 0.01$.

\begin{figure*}
\begin{subfigure}[b]{0.24\textwidth}
\centering
\includegraphics[width=\textwidth]{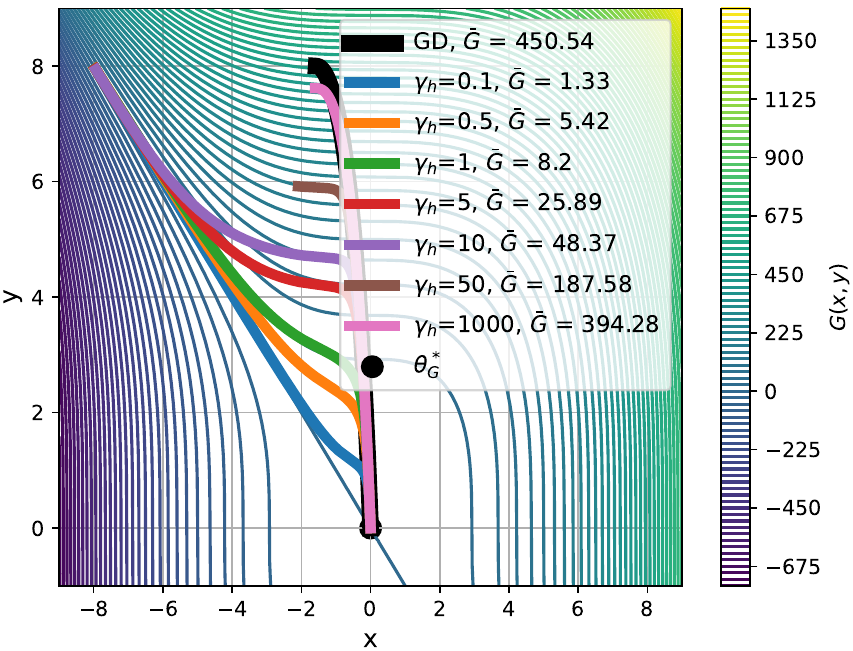}
\caption[]%
{{Parameter trajectories.}}    
\label{fig:cbfpa_gamma_test}
\end{subfigure}
\begin{subfigure}[b]{0.24\textwidth}
\centering
\includegraphics[width=\textwidth]{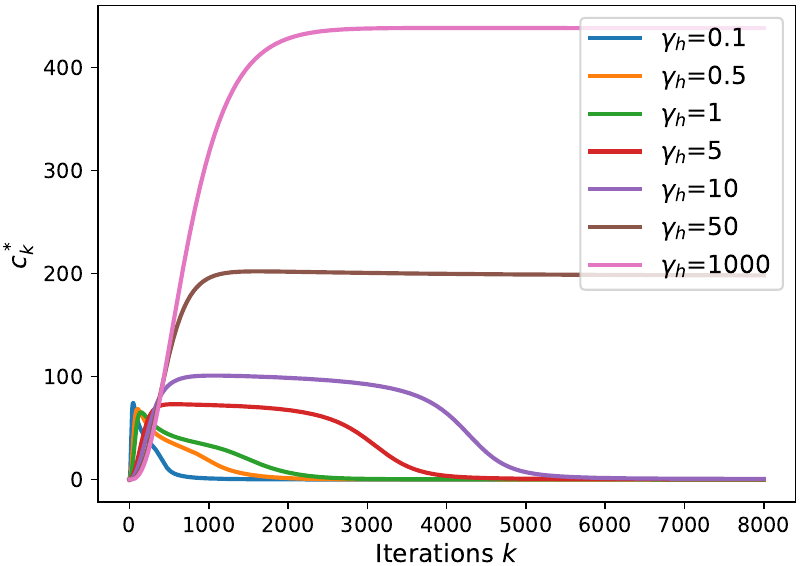}
\caption[]%
{{Evolution of $c_k^*$.}}
\label{fig:cbfpa_gamma_c_test}
\end{subfigure}
\begin{subfigure}[b]{0.24\textwidth}
\centering
\includegraphics[width=\textwidth]{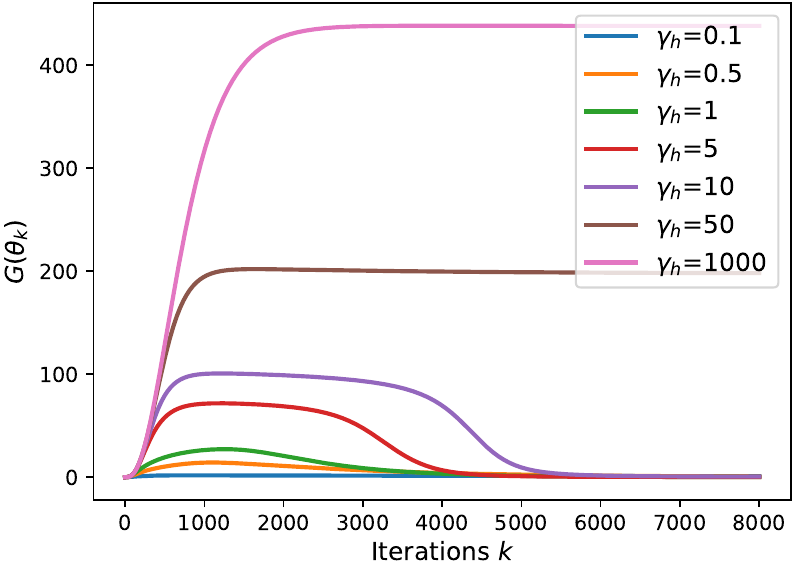}
\caption[]%
{{Original cost.}}    
\label{fig:cbfpa_gamma_G_test}
\end{subfigure}
\begin{subfigure}[b]{0.24\textwidth}
\centering
\includegraphics[width=\textwidth]{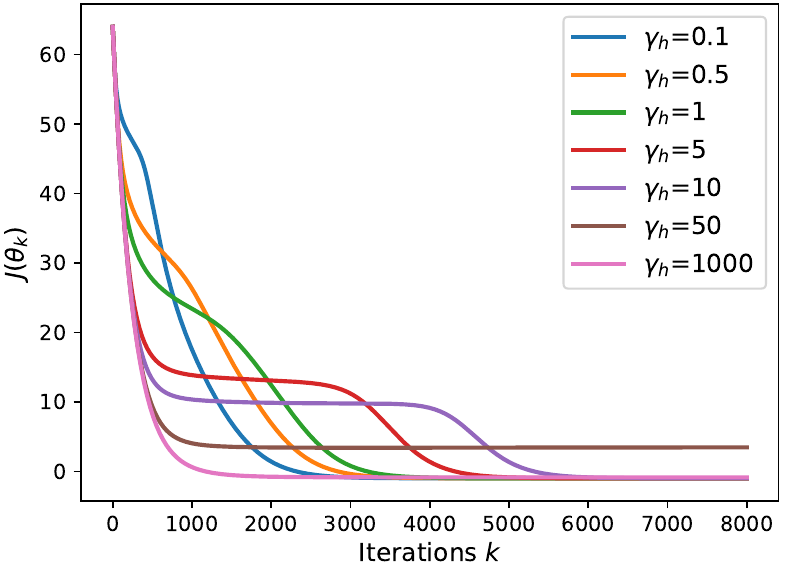}
\caption[]%
{{Additional cost.}}    
\label{fig:cbfpa_gamma_J_test}
\end{subfigure}
\caption{Proposed CBF-PA for different $\gamma_{\mathrm{h}}$, plotted on the contours of $G(x, y)$.} 
\label{fig:cbfpa_gamma}
\end{figure*}

\begin{figure*}
\begin{subfigure}[b]{0.24\textwidth}
\centering
\includegraphics[width=\textwidth]{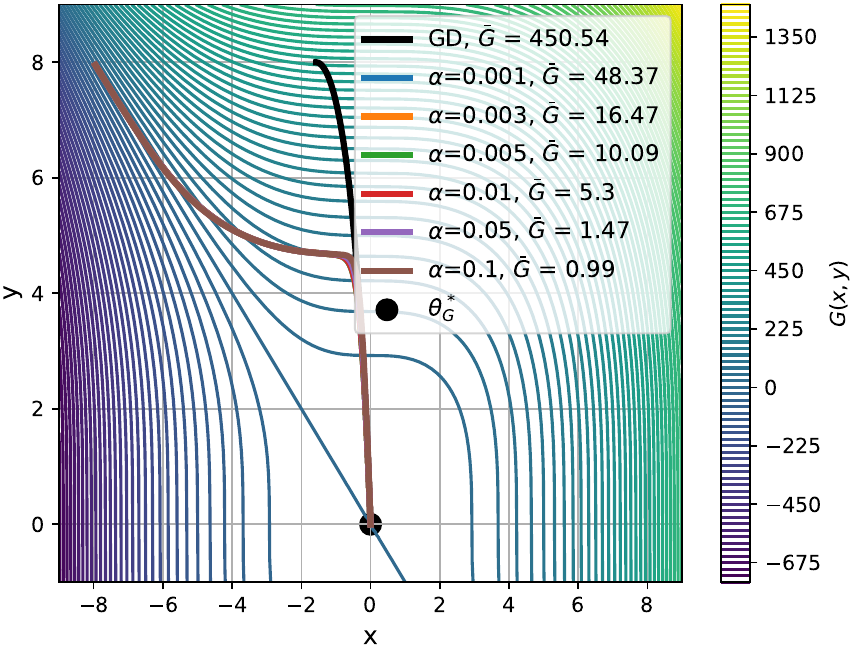}
\caption[]%
{{Parameter trajectories.}}    
\label{fig:cbfpa_alpha_test}
\end{subfigure}
\begin{subfigure}[b]{0.24\textwidth}
\centering
\includegraphics[width=\textwidth]{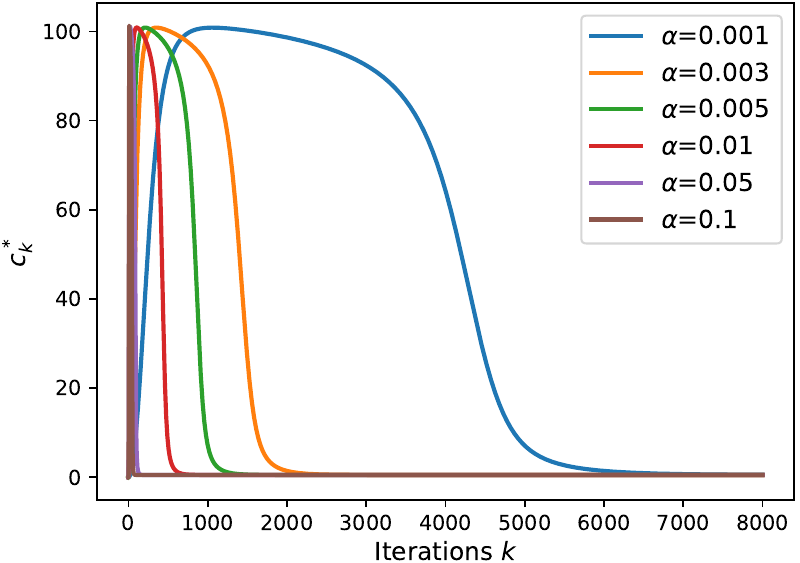}
\caption[]%
{{Evolution of $c_k^*$.}}
\label{fig:cbfpa_alpha_c_test}
\end{subfigure}
\begin{subfigure}[b]{0.24\textwidth}
\centering
\includegraphics[width=\textwidth]{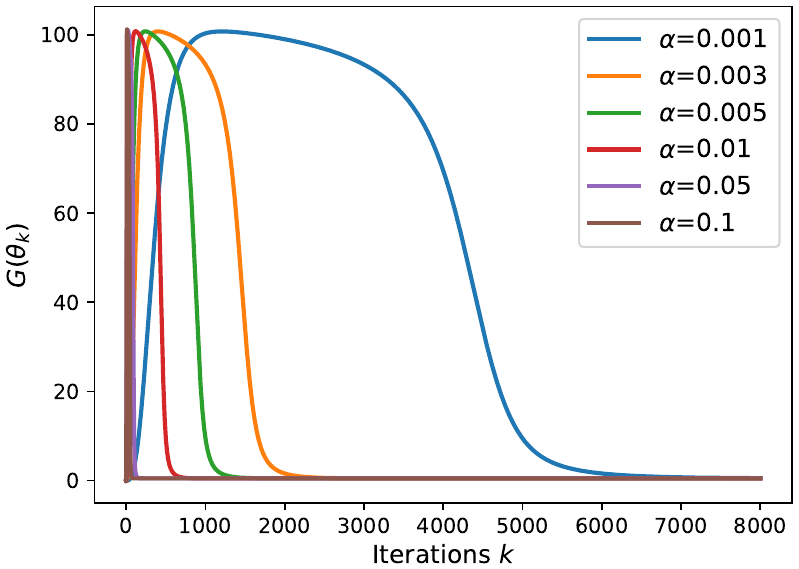}
\caption[]%
{{Original cost.}}
\label{fig:cbfpa_alpha_G_test}
\end{subfigure}
\begin{subfigure}[b]{0.24\textwidth}
\centering
\includegraphics[width=\textwidth]{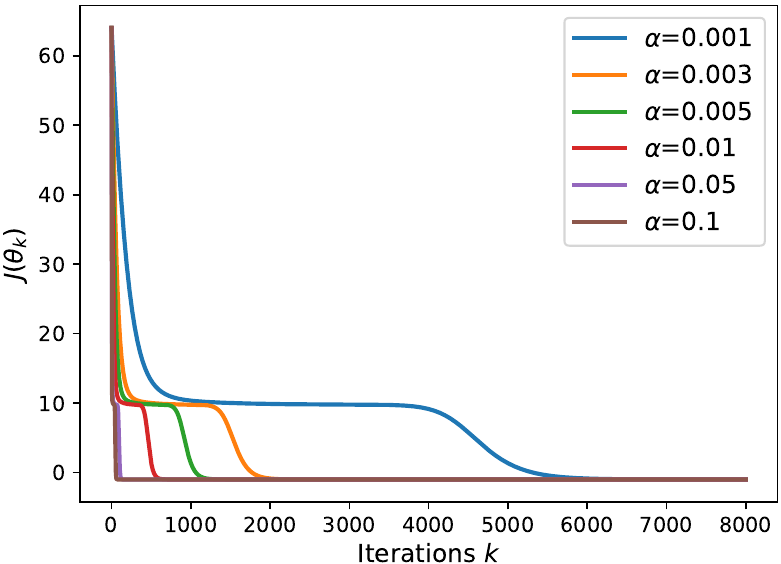}
\caption[]%
{{Additional cost.}}
\label{fig:cbfpa_alpha_J_test}
\end{subfigure}
\caption{Proposed CBF-PA for different $\alpha$, plotted on the contours $G(x, y)$.} 
\label{fig:cbfpa_alpha}
\end{figure*}

\textbf{Results Analysis.}
As shown in Figs. \ref{fig:cbfpa_gamma}-\ref{fig:cbfpa_alpha}, reducing $\gamma_{\mathrm{h}}$ from $10$ to $0.1$ accelerates the convergence of both $J$ and $G$. Conversely, when $\gamma_{\mathrm{h}} \geq 50$, CBF-PA tends to prioritize minimizing the additional cost $J$ over the original cost $G$. Furthermore, reducing the step size $\alpha$ leads to slower convergence while yielding parameter trajectories similar to those of the proposed method with larger $\alpha$.

\subsection{Cartpole Example}
The state of the Cartpole system is represented as $\boldsymbol{x}(t) = [p_{\mathrm{x}}(t), \dot{p}_{\mathrm{x}}(t), \theta(t), \dot{\theta}(t)]'$, where $-4.8 \leq p_{\mathrm{x}}(t) \leq 4.8$ denotes the cart's position, $\dot{p}_{\mathrm{x}}(t)$ represents the cart's velocity, $-0.418 \leq \theta(t)\leq 0.418$ is the pole's angle, and $\dot{\theta}(t)$ is the pole's angular velocity. The control input $-1 \leq \boldsymbol{u}(t) \leq 1$ is applied to the center of the cart to push the cart to move left or right. The dynamics of the Cartpole are given by:
\begin{equation} \label{eq_dyn_theta_cart}
\begin{split}
&\boldsymbol{x}(t+1) = \boldsymbol{x}(t) +  \dot{\boldsymbol{x}}(t) \Delta t, \\
&\Ddot{p}_{\mathrm{x}} = \frac{u + lm_{\mathrm{p}}\dot{\theta}^2\sin(\theta)}{m_{\mathrm{c}} + m_{\mathrm{p}}} - \frac{lm_{\mathrm{p}}\Ddot{\theta}\cos(\theta) }{m_{\mathrm{c}} + m_{\mathrm{p}}}, \\
&\Ddot{\theta} = \frac{g\sin(\theta)-\cos(\theta)(u + lm_{\mathrm{p}}\dot{\theta}^2\sin(\theta))/(m_{\mathrm{c}} + m_{\mathrm{p}})}{l(4/3 - m_{\mathrm{p}}\cos(\theta)^2/(m_{\mathrm{c}} + m_{\mathrm{p}}))},
\end{split}
\end{equation}
where $l$, $g$, $m_{\mathrm{c}}$, $m_{\mathrm{p}}$ denote the length of the pole, gravity, the mass of the cart, and the mass of the pole, respectively. 

\textbf{Setup.} We set the time interval $\Delta t = 0.05$ seconds for simulating the dynamics in \eqref{eq_dyn_theta_cart} to generate the training data. The control inputs are derived from the policy $\boldsymbol{u}(t) = \boldsymbol{\mu}(\boldsymbol{x}(t), \boldsymbol{\theta}_k^\mu) + W(t)$, where $W(t)$ represents exploration noise sampled from a normal distribution. The primary control objective for the Cartpole is to maintain the pole upright by appropriately moving the cart. For this, a predefined stage cost is defined as follows: \begin{equation}\label{eq_oricost_cart}
    \hat{\phi}(\theta(t)) = \begin{cases}
    -1, \quad &\text{if} -0.2095\leq \theta(t) \leq 0.2095, \\
     0, \quad &\text{else}. 
    \end{cases} 
\end{equation}
The predesigned optimal policy $\hat \mu$ and critic $\hat{Q}$ are achieved by running the existing DDPG algorithm. An additional objective is to drive the cart to the position $p_{\mathrm{x}}(t)=2$, for which we define the stage cost function as: \begin{equation}\label{eq_addcost_cart}
    \phi(p_{\mathrm{x}}(t), \dot{p}_{\mathrm{x}}(t)) = 0.1(p_{\mathrm{x}}(t)-2)^2 + 0.001\dot{p}_{\mathrm{x}}(t)^2.
\end{equation}
For comparison purposes, we define the stage cost for multi-objective RL as: $\phi_t^{MORL} = \hat{\phi}(\theta(t)) + w\phi(p_{\mathrm{x}}(t), \dot{p}_{\mathrm{x}}(t))$, and define the stage cost based on the concept of behavior cloning \cite{nair2018overcoming} from transfer learning as: \[\phi_t^{BC} = \phi(p_{\mathrm{x}}(t), \dot{p}_{\mathrm{x}}(t)) + w(\boldsymbol{\mu}(\boldsymbol{x}(t), \boldsymbol{\theta}^\mu) - \boldsymbol{\mu}(\boldsymbol{x}(t), \boldsymbol{\theta}^{\hat\mu *}))^2.\] Finally, all algorithms are trained for $200$ episodes. Each episode terminates either when the state constraint is violated or when the time step $t$ exceeds $200$. The initial state for each episode is randomly sampled from a uniform distribution within the range $[-0.05, -0.05, -0.05, -0.05]'$ to $[0.05, 0.05, 0.05, 0.05]'$. The proposed method is deployed to minimize the accumulated cost $\phi$ in \eqref{eq_addcost_cart}, while the DDPG algorithm is implemented to minimize $\phi_t^{BC}$ and $\phi_t^{MORL}$. All methods are initialized from the same predesigned policy $\boldsymbol{\mu}$ and critic $\hat{Q}$. To mitigate the effects of stochasticity arising from random initialization and training of DNNs, all experiments are repeated over $5$ independent trials.
\begin{figure}
\centering
\begin{subfigure}[b]{0.30\textwidth}
\includegraphics[width=\textwidth]{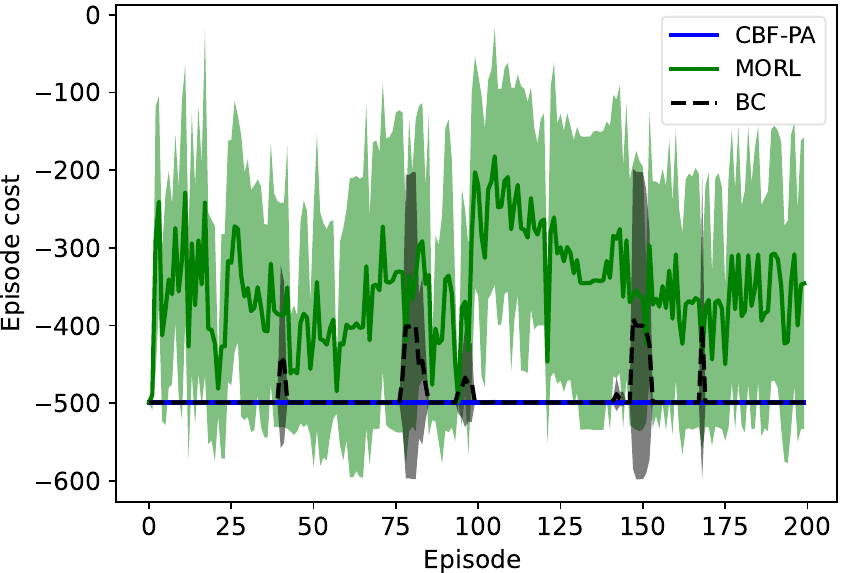}
\caption[]%
{{Cost of original task $\hat{\phi}(\theta(t))$.}}    
\label{fig:cart_ori}
\end{subfigure}
\begin{subfigure}[b]{0.30\textwidth}
\centering
\includegraphics[width=\textwidth]{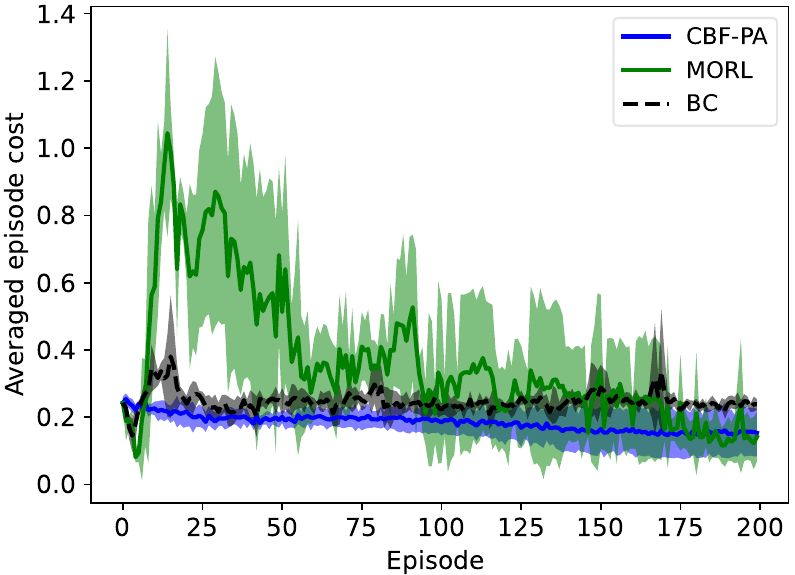}
\caption[]%
{{Cost of additional task $\phi(p_{\mathrm{x}}(t), \dot{p}_{\mathrm{x}}(t))$.}}
\label{fig:cart_add}
\end{subfigure}
\caption{Training cost for the Cartpole example, where Fig.~\ref{fig:cart_ori} denotes the episode cost refers to the cumulative original stage cost $\hat{\phi}$ in \eqref{eq_oricost_cart} accumulated over each episode, while Fig.~\ref{fig:cart_add} is the averaged episode cost represents the average additional stage cost $\phi$ in \eqref{eq_addcost_cart}, accounting for variations in initial positions. The solid line in the plots illustrates the mean value across $5$ experimental trials, and the shaded region indicates the standard deviation over these trials.} 
\label{fig:cart_train}
\end{figure}

\begin{figure}[ht]
\centering
\begin{subfigure}{0.75\linewidth}
\centering
\includegraphics[width=\linewidth]{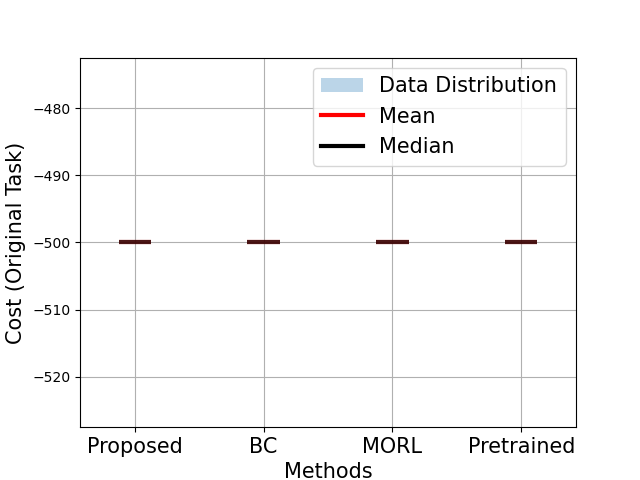}
\caption{Cost of original task.}
\label{fig:cart_cost:origin}
\end{subfigure}
\begin{subfigure}{0.70\linewidth}
\centering
\includegraphics[width=\linewidth]{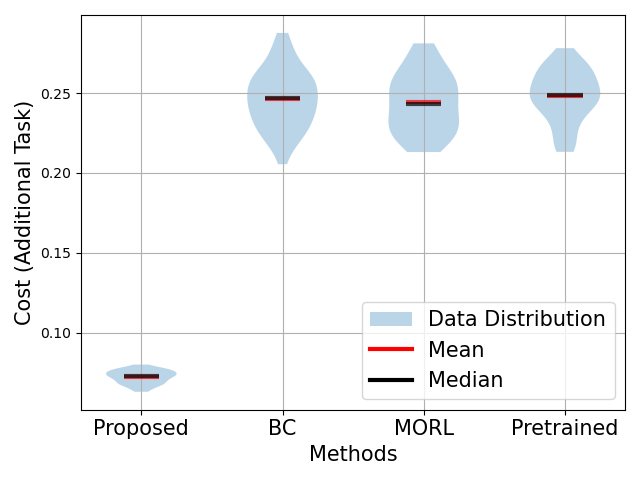}
\caption{Cost of additional task $\phi(p_{\mathrm{x}}(t), \dot{p}_{\mathrm{x}}(t))$.}
\label{fig:cart_cost:add}
\end{subfigure}
\caption{Cost of original and additional task for 50 test episodes with the Cartpole.}
\label{fig:cart_cost}
\end{figure}

\textbf{Results Analysis.} Fig. \ref{fig:cart_train} illustrates the training cost for all comparison algorithms. The proposed algorithm effectively preserves the original task, reaching a cost of $-499$ over the $200$ training epochs across $5$ trials. Moreover, it exhibits a faster convergence rate in minimizing the additional stage cost compared to the baseline methods. Fig.~\ref{fig:cart_cost} illustrates the performance of adapted policies across all evaluated algorithms over $50$ test episodes.
In Fig.~\ref{fig:cart_cost:origin}, all methods maintain optimal performance on the original task, successfully balancing the Cartpole for the $500$ time steps for each episode, indicating that no method compromises the original objective.
In contrast, Fig.~\ref{fig:cart_cost:add} highlights clear differences in performance on the additional task: the proposed algorithm significantly reduces the additional stage cost compared to the baseline methods.

To evaluate statistical significance, a one-way analysis of variance (ANOVA) \cite[Chapter~14]{lowry2014concepts} is conducted. The analysis reveals a statistically significant difference in the additional task cost across methods (p-value $<$ 0.05). Post-hoc comparisons using Tukey’s HSD test indicate the following relationship: Proposed algorithm $<$ Behavior Cloning $=$ Multi-objective RL $=$ Pretrained model. This result confirms that the proposed algorithm achieves the additional objective more effectively without sacrificing performance on the original task.

\subsection{Lunar Lander Example}
In this subsection, we evaluate the proposed CBF‑PA algorithm on the OpenAI Gym Lunar Lander environment \cite{brockman2016openai}. The system state $\boldsymbol{x}(t)\in\mathbb{R}^8$ comprises the horizontal and vertical positions, velocities, angular orientation, angular velocity, and ground-contact indicators of the lander and is subject to $[-1.5, -1.5, -5.0, -5.0, -3.14, -5.0, 0, 0]'\leq\boldsymbol{x}(t)\leq [1.5, 1.5, 5.0, 5.0, 3.14, 5.0, 1.0, 1.0]'$. The control input $\boldsymbol{u}(t)\in\mathbb{R}^2$, which commands the main and lateral thrusters, is similarly bounded by $[-1, -1]'\leq\boldsymbol{u}(t)\leq[1,1]'$. For a detailed description of the underlying dynamics, see \cite{brockman2016openai}.

\textbf{Setup.} As in the Cartpole example, the system state-input data pairs are generated using the policy $\boldsymbol{u}(t) = \boldsymbol{\mu}(\boldsymbol{x}(t), \boldsymbol{\theta}_k^\mu) + W(t)$, where $W(t)$ represents exploration noise drawn from a normal distribution. Each episode begins with the lunar lander spawned at a fixed initial pose, subject to a random perturbation at its center of mass, and terminates when either a state constraint is violated or the time index $t$ exceeds $200$. The primary control objective is to achieve a soft touchdown at the target pad located at $(x, y) = (0,0)$. The underlying predefined stage cost $\hat{\phi}(\boldsymbol{x}(t), \boldsymbol{u}(t))$ are adopted from \cite{brockman2016openai}. Both the predesigned policy $\boldsymbol{\mu}$ and critic $\hat{Q}$ are obtained by running the DDPG algorithm. To evaluate the proposed CBF‑PA framework, an additional task to minimize energy consumption during the landing phase is defined as: \begin{equation}\label{eq_addcost_lunar}
    \phi(\boldsymbol{u}(t)) = \boldsymbol{u}(t)'\boldsymbol{u}(t).
\end{equation} 
For comparative analysis, we define the stage cost for multi-objective RL as: $\phi_t^{MORL} = \hat{\phi}(\boldsymbol{x}(t), \boldsymbol{u}(t)) + w\phi(\boldsymbol{u}(t))$, and for behavior cloning \cite{nair2018overcoming} as $\phi_t^{BC} = \phi(\boldsymbol{u}(t)) + w\|\boldsymbol{\mu}(\boldsymbol{x}(t), \boldsymbol{\theta}^\mu) - \boldsymbol{\mu}(\boldsymbol{x}(t), \boldsymbol{\theta}^{\hat\mu *})\|^2$.
Each algorithm was trained for $200$ episodes and evaluated across $5$ independent trials.
\begin{figure}
\centering
\begin{subfigure}[b]{0.35\textwidth}
\centering
\includegraphics[width=\textwidth]{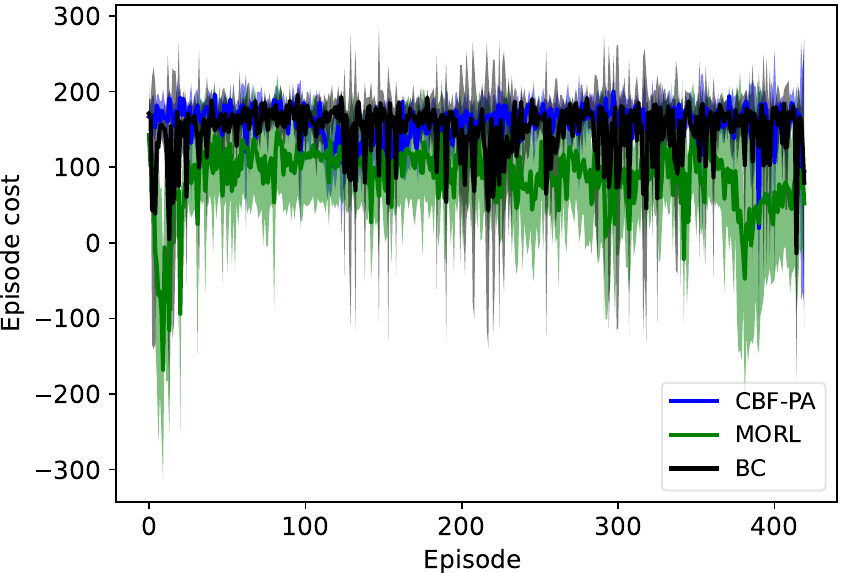}
\caption[]%
{{Cost of original task.}}    
\label{fig:train_ori_lander}
\end{subfigure}
\begin{subfigure}[b]{0.35\textwidth}
\centering
\includegraphics[width=\textwidth]{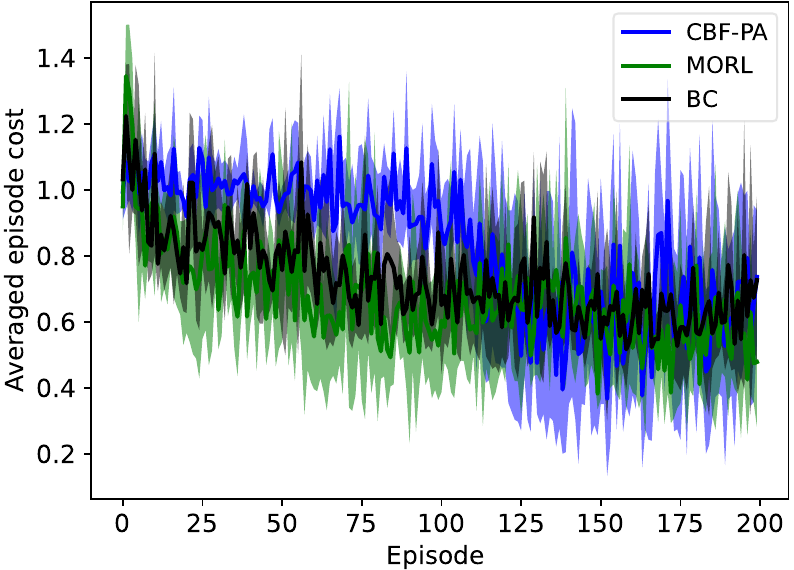}
\caption[]%
{{Cost of additional task $\phi(\boldsymbol{u}(t))$.}}
\label{fig:train_add_lander}
\end{subfigure}
\caption{Training plots for lunar lander example, where Fig.~\ref{fig:train_ori_lander} denotes the episode rewards, which refers to the original rewards accumulated over each episode, while Fig.~\ref{fig:train_add_lander} denotes the averaged episode cost, representing the average additional stage cost $\phi$ in \eqref{eq_addcost_lunar}, accounting for variations in initial positions. The solid line in the plots illustrates the mean value across $5$ experimental trials, and the shaded region indicates the standard deviation over these trials.} 
\label{fig:lunar_train}
\end{figure}

\begin{figure}
\centering
\begin{subfigure}{0.8\linewidth}
\centering
\includegraphics[width=\linewidth]{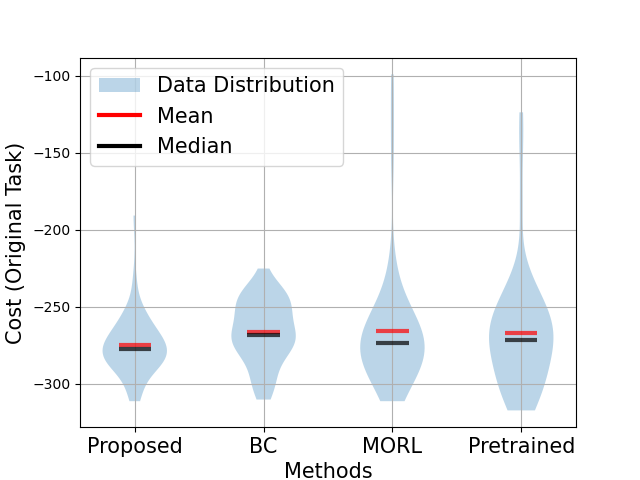}
\caption{Cost of original task.}
\label{fig:lunar_cost:origin}
\end{subfigure}
\begin{subfigure}{0.73\linewidth}
\centering
\includegraphics[width=\linewidth]{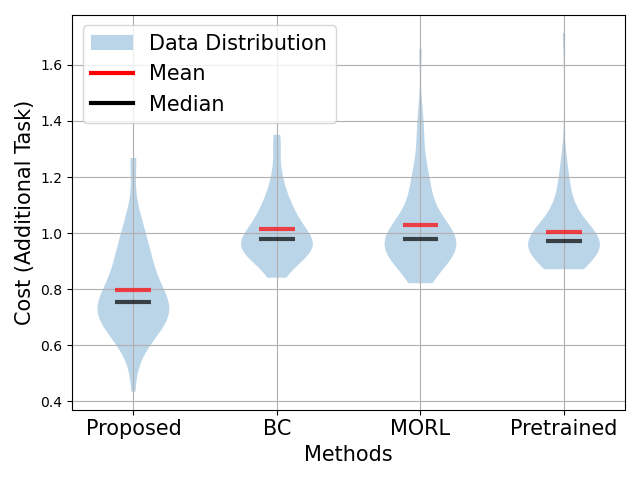}
\caption{Cost of additional task $\phi(\boldsymbol{u}(t))$.}
\label{fig:lunar_cost:add}
\end{subfigure}
\caption{Cost of original and additional tasks for 50 test episodes with the lunar lander.}
\label{fig:lunar_cost}
\end{figure}

\textbf{Results Analysis.} Fig. \ref{fig:lunar_train} showcases the training process for all comparison algorithms. The proposed algorithm successfully maintains the original task over $400$ training episodes across $5$ trials. Moreover, it demonstrates faster convergence, reducing the additional stage cost within $200$ episodes, surpassing the baseline methods in efficiency. Fig.~\ref{fig:lunar_cost} presents the performance of the adapted policies across all evaluated algorithms over $50$ test episodes.
In Fig.~\ref{fig:lunar_cost:origin}, all methods successfully maintain the objective on the original task, enabling the lunar lander to reach the designated landing position.
To assess statistical differences, an ANOVA \cite[Chapter~14]{lowry2014concepts} was conducted, which finds no statistically significant differences in the original task cost among methods (p-value $>0.05$).

In contrast, Fig.~\ref{fig:lunar_cost:add} reveals notable differences in the additional task performance. The proposed algorithm achieves a substantially lower additional stage cost, effectively minimizing energy consumption during the landing process compared to baseline methods.
An ANOVA confirms that the differences in additional task cost are statistically significant (p-value $< 0.05$). Post-hoc analysis using Tukey’s HSD test further identifies the following relationship: Proposed algorithm $<$ Behavior Cloning $=$ Multi-objective RL $=$ Pretrained model.
These results demonstrate that the proposed algorithm enhances performance on the additional objective with an acceptable degradation in performance on the original task.

\section{Real-World Demonstrations}\label{exp_dog}
\begin{figure}
\centering
\includegraphics[width=0.30\linewidth]{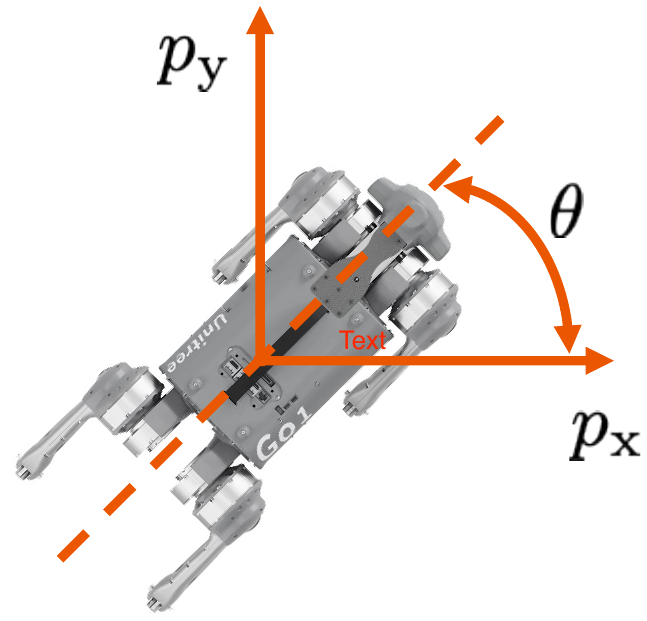}
\caption{High-level quadrupedal robot kinematics definition.}
\label{fig:dog_intro}
\end{figure}

In this section, we demonstrate the performance of the proposed CBF-PA  algorithm on a quadrupedal robot platform, highlighting its effectiveness in path planning adaptation scenarios. The kinematic dynamics of the quadrupedal robot are modeled as a unicycle model, described as:
\begin{equation}
\boldsymbol{\dot{x}} = \matt{\dot{p}_{\mathrm{x}} \\ \dot{p}_{\mathrm{y}} \\ \dot{\theta} } = \matt{u_{\mathrm{v}} \cos(\theta) \\ u_{\mathrm{v}} \sin(\theta) \\ u_\omega },
\end{equation}
where the system state and control input at any time $t$ are defined as $\boldsymbol{x}(t) = \matt{p_{\mathrm{x}}(t),p_{\mathrm{y}}(t),\theta(t)}' \in \mathbb{R}^3$ and 
$\boldsymbol{u}(t) = \matt{u_{\mathrm{v}}(t), u_\omega(t)}' \in \mathbb{R}^2$, respectively.
Here, $-2.4 \text{m} \leq p_{\mathrm{x}}\leq 2.4 \text{m}, -1.8 \text{m}\leq p_{\mathrm{y}}\leq 1.6 \text{m}, -\pi\leq\theta\leq \pi$ are the positions on the x-axis and y-axis, and yaw angle, respectively.
$-1 \text{ m/s} \leq u_{\mathrm{v}}\leq 1 \text{ m/s}$ and $-3 \text{ rad/s} \leq u_{\omega}\leq 3 \text{ rad/s}$ are the linear velocity and yaw angular velocity of the robot, respectively. The initial state is uniformly generated between $[-1.9,-0.2,-0.2]'$ and $[1.5,0.2,0.2]'$. An illustration of this unicycle model is shown in Fig. \ref{fig:dog_intro}.
\subsection{Policy Adaptation to a Different Target Position} \label{subsec:task_switch}
In this experiment, we apply the proposed method to adapt a pretrained policy that enables the robot to avoid obstacles while reaching a specified target position.
\begin{figure*}
\subfloat[Time-lapse trajectory with pretrained policy.]
{\label{fig:set1:pre} \includegraphics[width=0.32\linewidth]{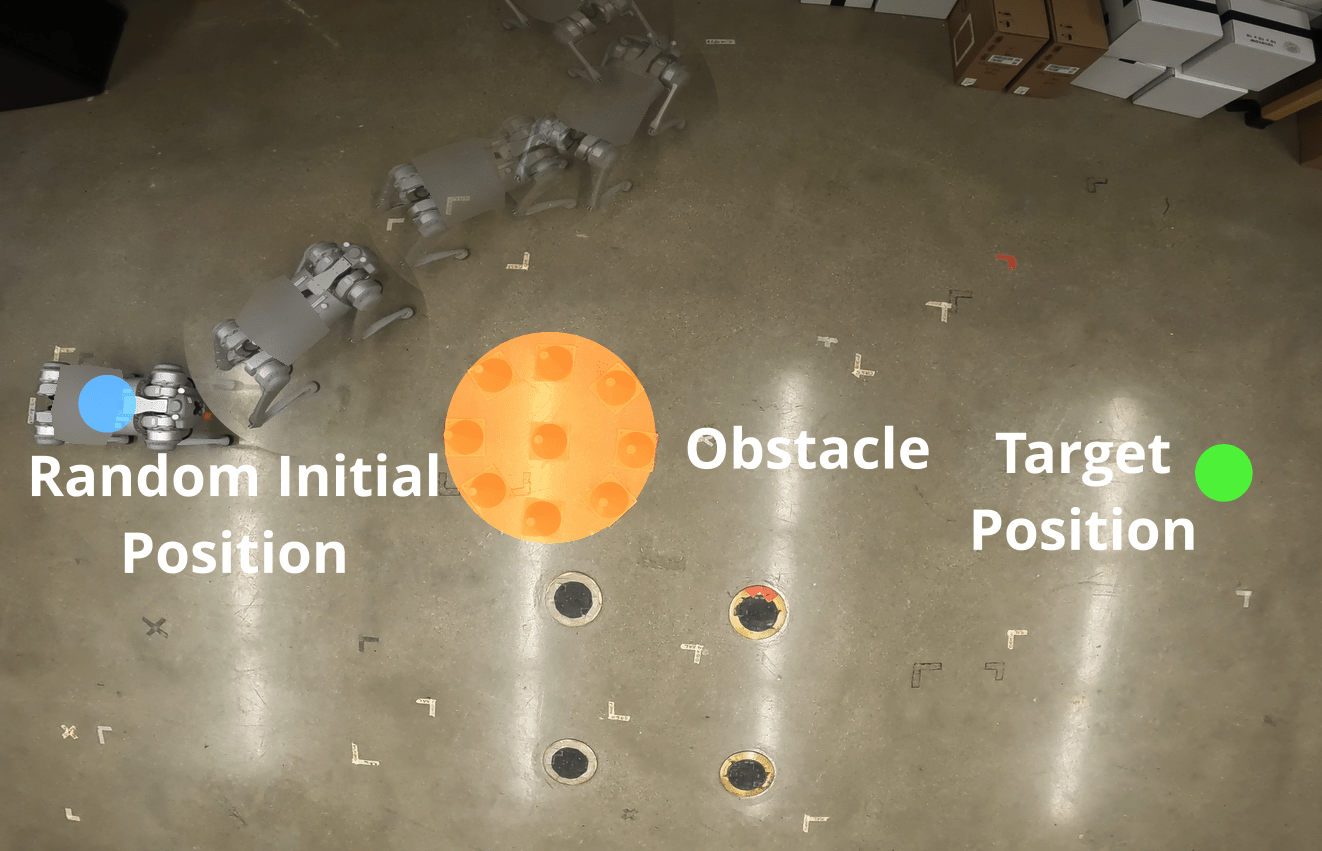}}
\hfill
\subfloat[Time-lapse trajectory with policy during adaptation training at episode 90.]
{\label{fig:set1:middle} \includegraphics[width=0.32\linewidth]{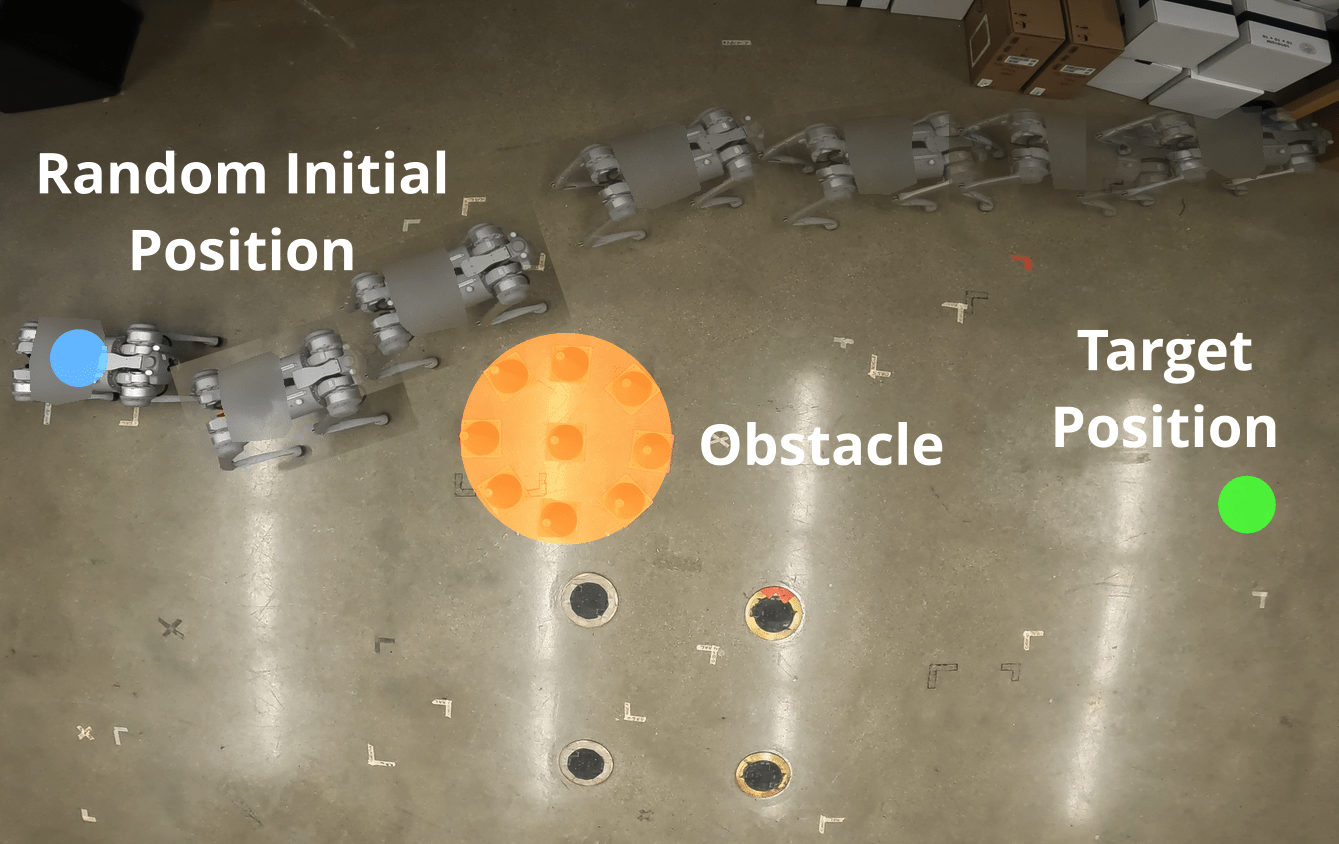}}
\hfill
\subfloat[Time-lapse trajectory with adapted policy.]
{\label{fig:set1:final_01} \includegraphics[width=0.32\linewidth]{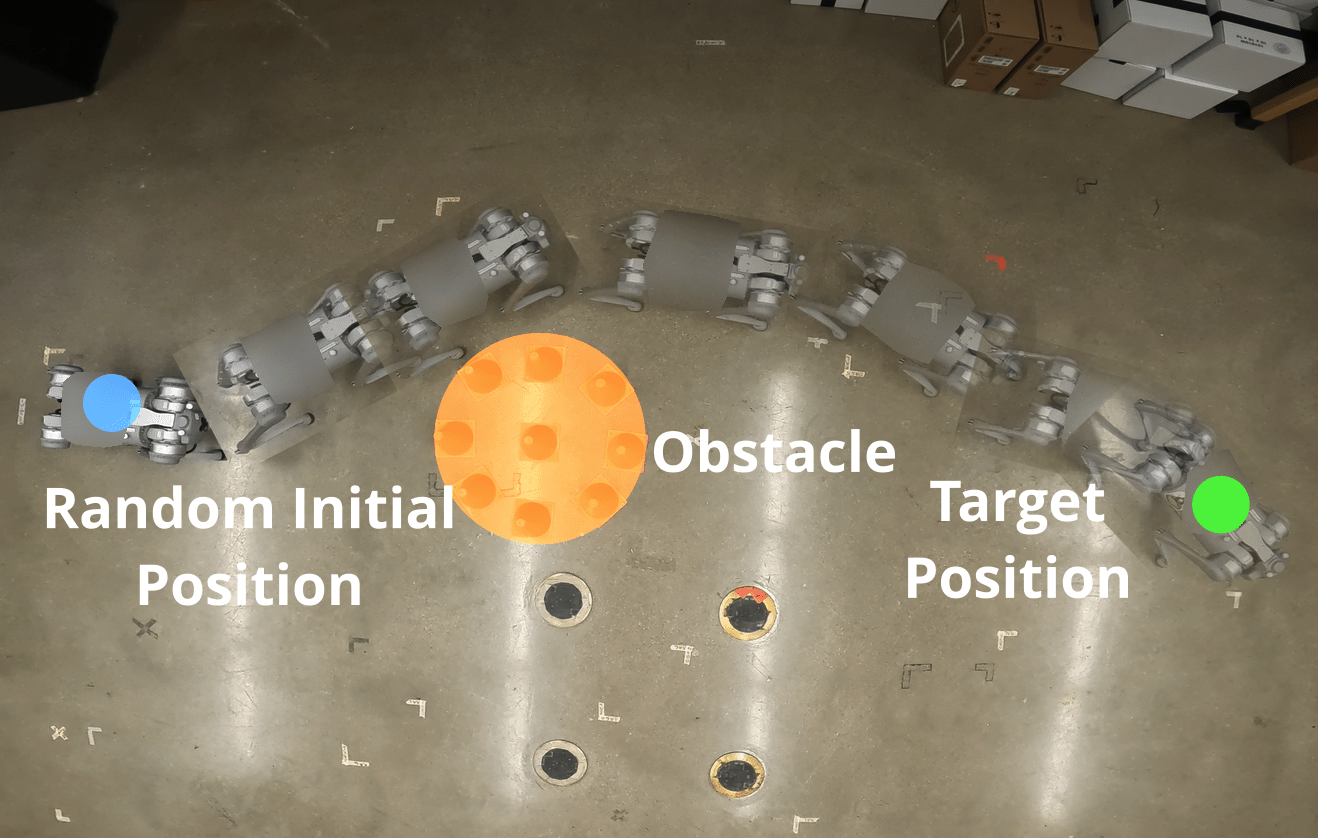}}
\hfill
\subfloat[Actual trajectory with pretrained policy.]
{\label{fig:set1:pre_traj} \includegraphics[width=0.32\linewidth]{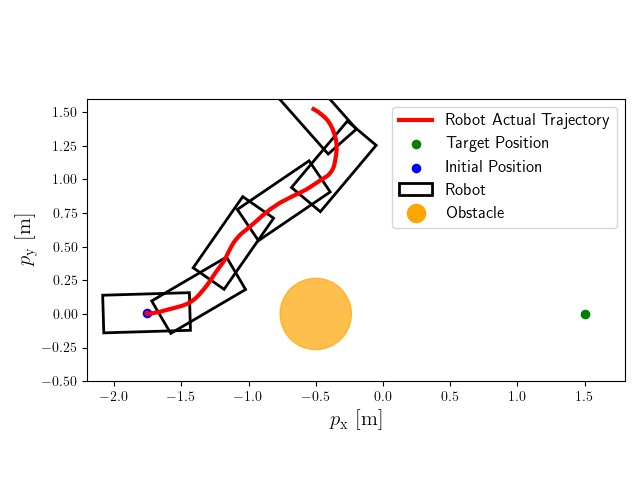}}
\hfill
\subfloat[Actual trajectory with policy during adaptation training at episode 90.]
{\label{fig:set1:middle_traj} \includegraphics[width=0.32\linewidth]{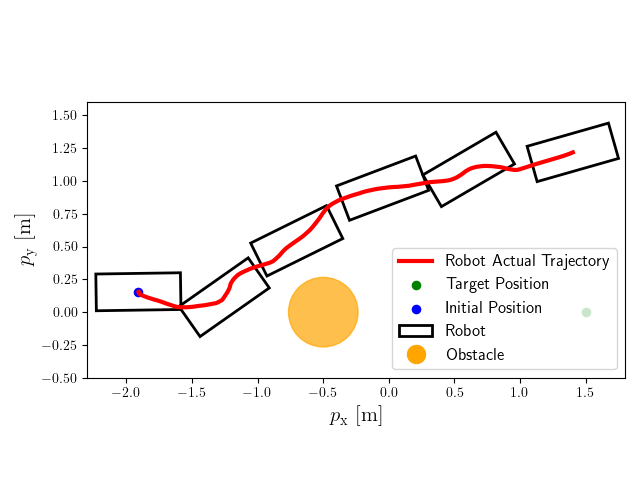}}
\hfill
\subfloat[Actual trajectory with adapted policy.]
{\label{fig:set1:final_01_traj} \includegraphics[width=0.32\linewidth]{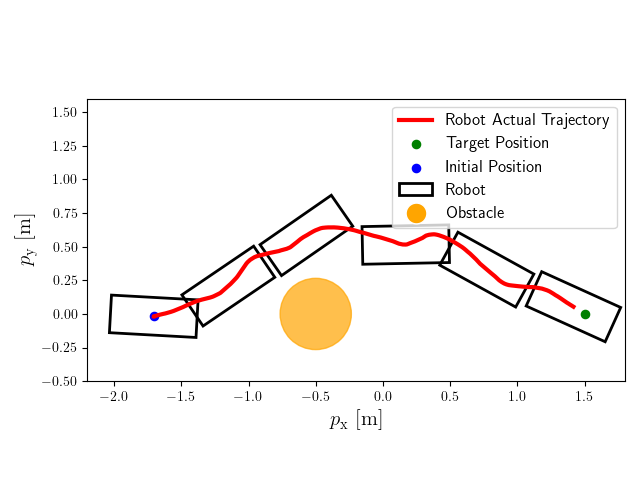}}
\caption{Time-lapse and actual position trajectories of the robot using the pretrained policy, the policy during adaptation training at episode 90, and the adapted policy. The pretrained policy is designed for the original task of obstacle avoidance. The adapted policy enables the robot to reach a target position (an additional task) while still avoiding collisions (the original task).} \label{fig:set1}
\end{figure*}

\begin{figure*}
\subfloat[Trajectories of the pretrained policy.]
{\label{fig:traj_pretrain} \includegraphics[width=0.32\linewidth]{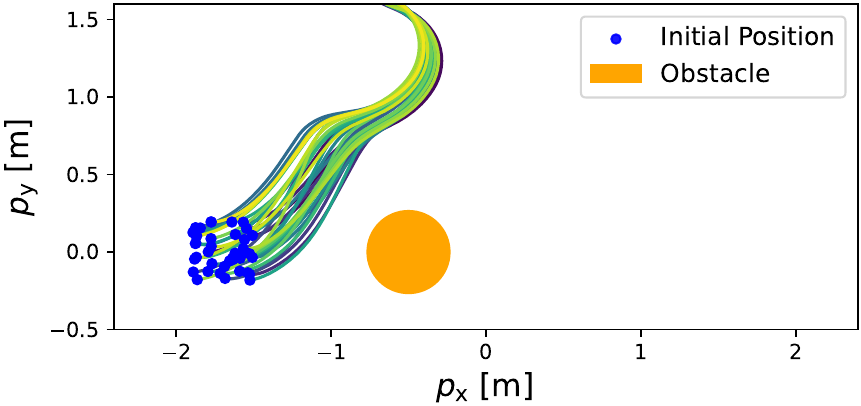}}
\hfill
\subfloat[Trajectories observed during adaptation.]
{\label{fig:traj_training} \includegraphics[width=0.32\linewidth]{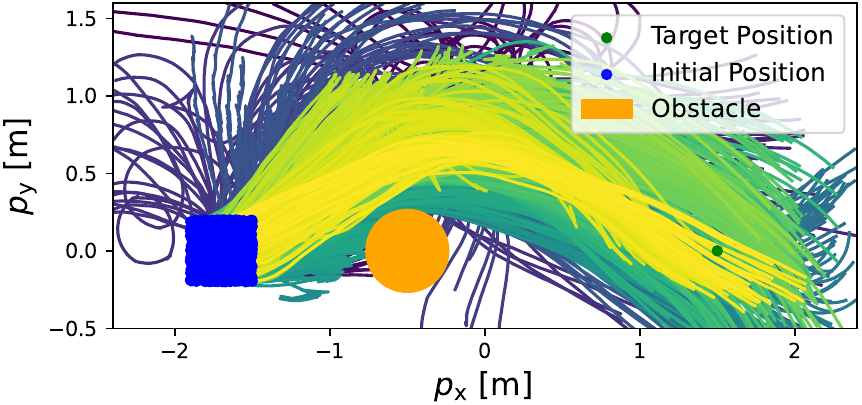}}
\hfill
\subfloat[Trajectories after adaptation.]
{\label{fig:traj_aftertrain} \includegraphics[width=0.32\linewidth]{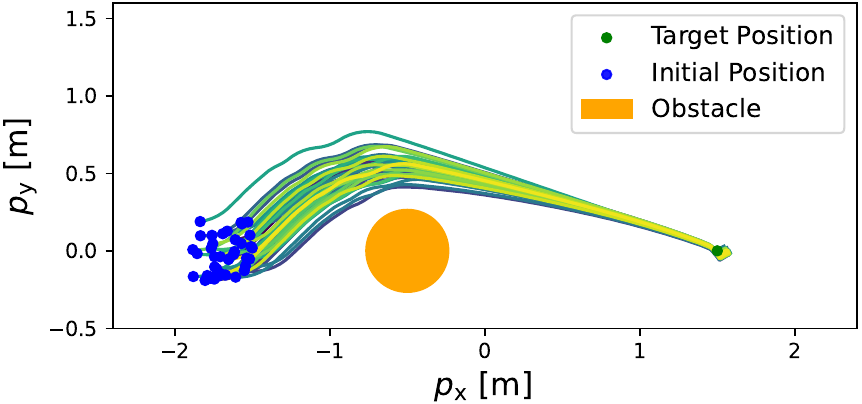}}
\caption{Demonstration trajectories. Colors indicate training epochs, where lighter colors correspond to later stages in training.} 
\label{fig:task1}
\end{figure*}

\textbf{Experiment Setup.} As shown in Fig. \ref{fig:set1:pre}, the original task of the quadrupedal robot is to avoid collision with the obstacle placed in the center with $p_{\mathrm{x}} = -0.5 \text{m}, p_{\mathrm{y}}=0 \text{m}$ and radius $0.265 \text{m}$,  for which we define the following predefined stage cost: \begin{equation}\label{eq_oricost_dog}
    \hat{\phi}(\boldsymbol{x}(t)) = \begin{cases}
    -(p_{\mathrm{x}}^2(t) + p_{\mathrm{y}}^2(t)) + 100, \quad &\text{if collision}, \\
     -(p_{\mathrm{x}}^2(t) + p_{\mathrm{y}}^2(t)), \quad &\text{else}, 
    \end{cases} 
\end{equation}
where the collision means $((p_{\mathrm{x}}(t) + 0.5)^2 + p_{\mathrm{y}}^2(t))^{\frac{1}{2}}\leq 0.595$ m. The stage cost $\hat{\phi}(\boldsymbol{x}(t))$ in \eqref{eq_oricost_dog} is designed to give the robot a penalty when the collision occurs and to encourage the robot to move away from the position $p_{\mathrm{x}} = 0, p_{\mathrm{y}}=0$. A pretrained policy $\boldsymbol{\mu}(\boldsymbol{x}(t), \boldsymbol{\theta}^{\hat \mu})$ and its corresponding critic $\hat Q(\boldsymbol{x}(t), \boldsymbol{u}(t), \boldsymbol{\theta}^{\hat Q})$ are obtained using the DDPG algorithm. A demonstration of trajectories generated by $\boldsymbol{\mu}$ is shown in Fig. \ref{fig:traj_pretrain}.
The additional task is to reach the target position $p_{\mathrm{x}} = 1.5 \text{m}, p_{\mathrm{y}}=0 \text{m}$, as shown in Fig. \ref{fig:set1:pre}. To achieve this, we define the following additional stage cost function:
\begin{equation}\label{eq_t1_add}
\phi(\boldsymbol{x}(t)) = \sqrt{(p_{\mathrm{x}}(t)-1.5)^2 + p_{\mathrm{y}}^2(t)}.
\end{equation}
The proposed algorithm is subsequently employed to adapt the original control policy to minimize the cost function in  \eqref{eq_t1_add}, while ensuring that the adapted policy continues to satisfy the original task constraints. The demonstration trajectories generated by the adapted policy $\boldsymbol{\mu}(\boldsymbol{x}(t),\boldsymbol{\theta}^{\mu})$ illustrated in Fig. \ref{fig:traj_aftertrain}, and the corresponding trajectories observed during the training process are presented in Fig. \ref{fig:traj_training}. This setup illustrates how the proposed algorithm can effectively adapt a pretrained policy to accommodate an additional objective, provided that the optimal solution sets for the two tasks have a non-empty intersection.

\begin{figure}
\centering
\begin{subfigure}{0.8\linewidth}
\centering
\includegraphics[width=\linewidth]{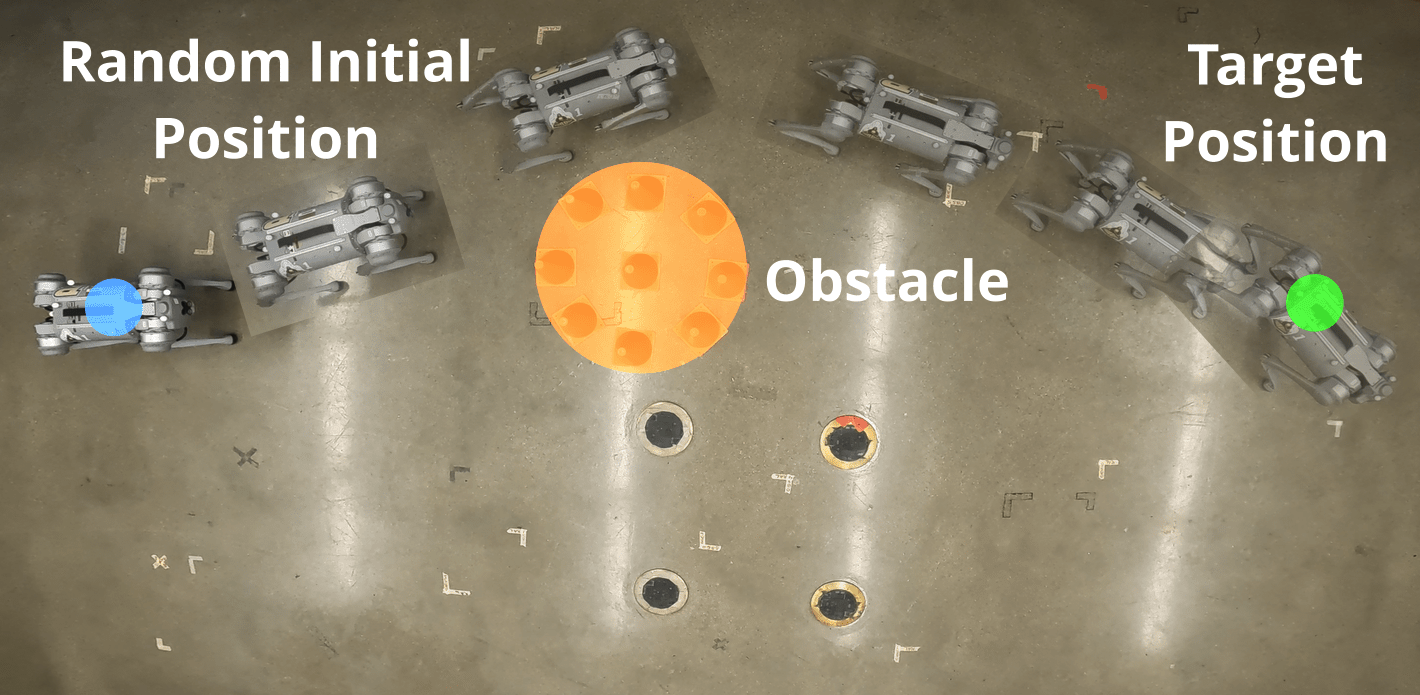}
\caption{Time-lapse trajectory with adapted policy.}
\label{fig:set2:pic}
\end{subfigure}
\begin{subfigure}{0.85\linewidth}
\centering
\includegraphics[width=\linewidth]{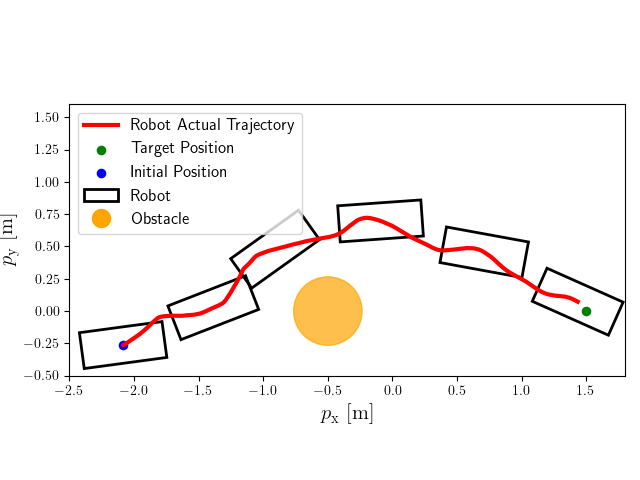}
\caption{Actual trajectory with adapted policy.}
\label{fig:set2:traj}
\end{subfigure}
\caption{Time-lapse and actual position trajectories of the robot using the same adapted policy with a different initial position. The adapted policy enables the robot to reach a target position (additional task) while still avoiding collisions (original task).}
\label{fig:set2}
\end{figure}

\textbf{Results Analysis.} As shown in Fig. \ref{fig:set1:pre}, the robot initially executes the pretrained policy, which was developed solely for obstacle avoidance. While the robot successfully avoids collisions, it fails to reach the target position since goal-reaching was not included in the original task. The actual trajectory of the robot, recorded via a motion capture system, is depicted in Fig. \ref{fig:set1:pre_traj}. Subsequently, as illustrated in Fig. \ref{fig:set1:middle}, the robot still does not reach the goal at the $90$-th episode of the adaptation training due to the incomplete adaptation process. Nevertheless, the proposed algorithm effectively utilizes information from both the original policy and the additional goal-reaching task to guide the adaptation. The corresponding trajectory is shown in Fig. \ref{fig:set1:middle_traj}. After completing the adaptation training, the robot executes the fully adapted policy. As demonstrated in Fig. \ref{fig:set1:final_01}, the robot successfully reaches the goal without any collisions, thus fulfilling both the original (collision avoidance) and additional (goal-reaching) task objectives. The trajectory for this successful trial is also shown in Fig. \ref{fig:set1:middle_traj}. Furthermore, to assess the generalizability of the adapted policy, an additional trial is conducted from a different initial position. As shown in Fig.~\ref{fig:set2}, the robot reaches the goal without collisions, demonstrating the effectiveness of the adapted policy in meeting both original and additional task objectives. Details are available in the supplementary video.

\subsection{Policy Adaptation to an Additional Obstacle}
In this experiment, we aim to utilize the proposed method to adapt the previously learned policy, which enables the robot to avoid obstacles and reach a specified goal, to additionally account for a new obstacle introduced into the environment.
\begin{figure}
\centering
\begin{subfigure}{0.8\linewidth}
\centering
\includegraphics[width=\linewidth]{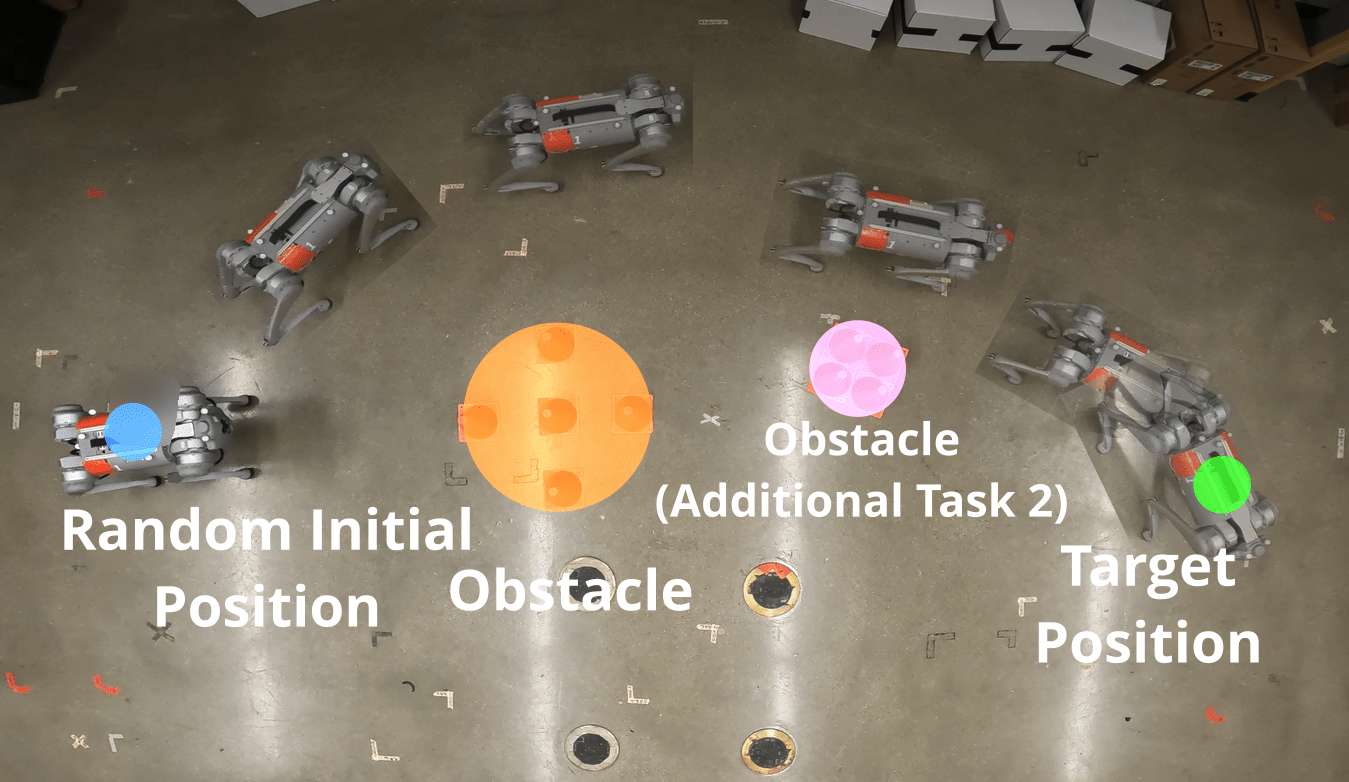}
\caption{Time-lapse trajectory with second adapted policy.}
\label{fig:add:1}
\end{subfigure}
\begin{subfigure}{0.85\linewidth}
\centering
\includegraphics[width=\linewidth]{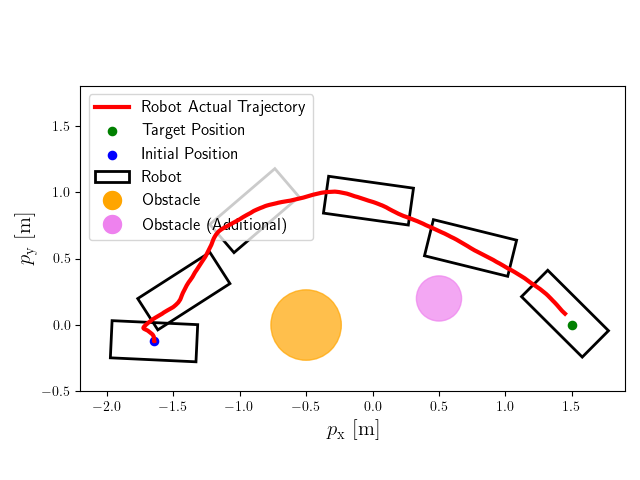}
\caption{Actual trajectory with second adapted policy.}
\label{fig:add:1_traj}
\end{subfigure}
\caption{Time-lapse and actual position trajectories of the robot using the first adapted policy as the new pretrained policy for the new original task. The new pretrained policy is designed for the new original task of orange obstacle avoidance and goal reaching. The second adapted policy enables the robot to reach a target position while still avoiding collisions with original (orange) and additional (violet) obstacles.}
\label{fig:add}
\end{figure}


\textbf{Experiment Setup.}  As shown in Fig. \ref{fig:add}, under the same setup as in Section \ref{subsec:task_switch}, the new original task for the quadrupedal robot is to avoid collision with the orange obstacle and reach the target position using the resulting policy from Section \ref{subsec:task_switch}. 

\begin{figure*}
\subfloat[Trajectories with no adaptation.]
{\label{fig:traj_notraining} \includegraphics[width=0.32\linewidth]{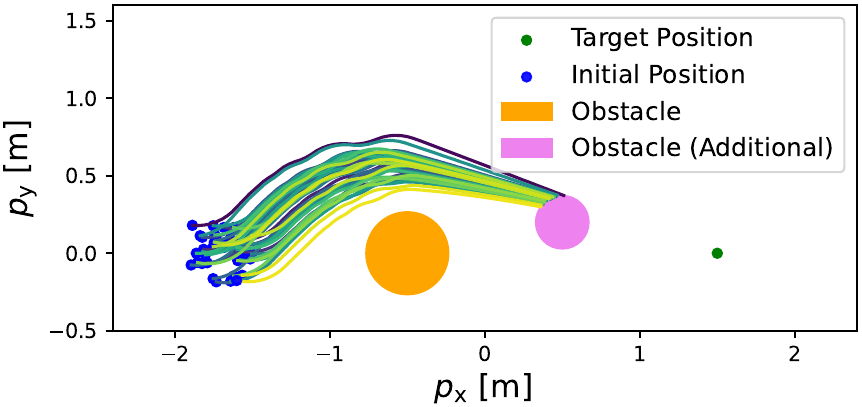}}
\subfloat[Trajectories observed during adaptation.]
{\label{fig:traj_training2} \includegraphics[width=0.32\linewidth]{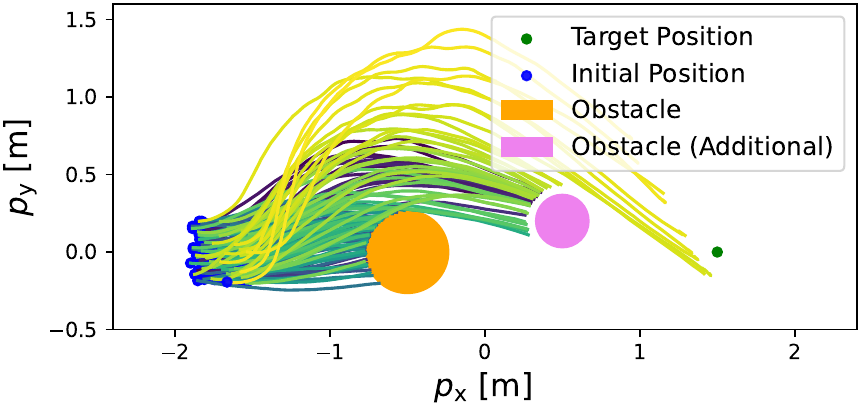}}
\subfloat[Trajectories after adaptation.]
{\label{fig:traj_aftertrain2} \includegraphics[width=0.32\linewidth]{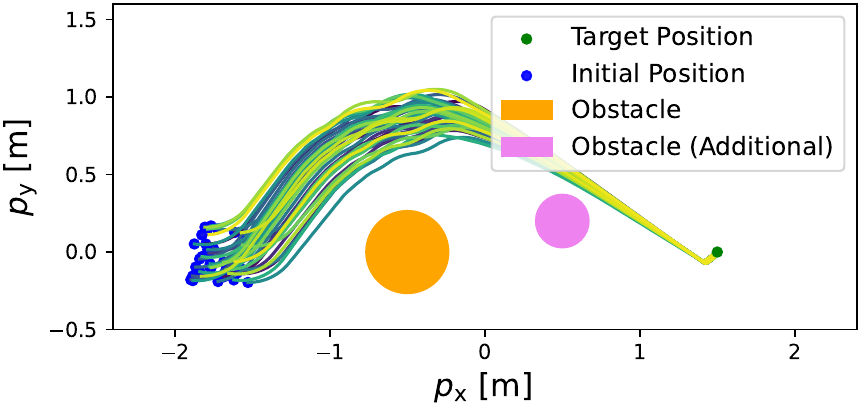}}
\caption{Demonstration trajectories. Colors indicate training epochs, where lighter colors correspond to later stages in training.} 
\label{fig:task2}
\end{figure*}

The additional task is to avoid collision with a newly introduced obstacle located at $p_{\mathrm{x}} = 0.5 \text{m}, p_{\mathrm{y}} = 0.2\text{m}$ with a radius of $0.17 \text{m}$, as shown in Fig. \ref{fig:add} and marked in violet. Notably, this obstacle is deliberately positioned on the planned trajectories generated by the resulting policy from Section \ref{subsec:task_switch}, as illustrated in Fig. \ref{fig:traj_notraining}. To incorporate this additional task, we define the following stage cost function:
\begin{equation}\label{eq_oricost_doL_add}
\phi(\boldsymbol{x}(t)) = \begin{cases}
\sqrt{(p_{\mathrm{x}}(t)-1.5)^2 + p_{\mathrm{y}}^2(t)} + 10, \quad &\text{if collision}, \\
 \sqrt{(p_{\mathrm{x}}(t)-1.5)^2 + p_{\mathrm{y}}^2(t)}, \quad &\text{else}, 
\end{cases} 
\end{equation}
where the collision here means $((p_{\mathrm{x}}(t) - 0.5)^2 + (p_{\mathrm{y}}(t) - 0.2)^2)^{\frac{1}{2}}\leq 0.42 \text{m}$. The proposed algorithm is subsequently utilized to further adapt the initially modified control policy from Section \ref{subsec:task_switch}, enabling the robot to avoid the newly introduced obstacle. The adaptation process is illustrated through the training trajectories shown in Fig. \ref{fig:traj_training2}, while the trajectories generated by the final adapted policy are presented in Fig. \ref{fig:traj_aftertrain2}.

\textbf{Results Analysis.} Fig. \ref{fig:add} presents the experimental results. After completing training, the robot executes the second adapted policy. As shown in Fig. \ref{fig:add:1}, the robot successfully reaches the target position without any collisions, satisfying both the original task (collision avoidance for the orange obstacle and goal reaching) and the additional task (collision avoidance for the violet obstacle). The collision-free robot trajectory is illustrated in Fig. \ref{fig:add:1_traj}.
Multiple trials with different initial positions, available in the supplementary video, were conducted to evaluate the generalizability. These experiments also highlight that the proposed algorithm supports sequential policy adaptation to multiple additional tasks.


\section{Concluding Remarks}\label{Conc}
In this paper, we have presented a policy adaptation algorithm, termed CBF-PA, for adjusting the parameters of a predesigned policy to minimize an additional task while simultaneously preserving the near-optimal performance of the original task. The key contributions of this work are: (i) formulating the policy adaptation problem as a constrained optimization problem, and (ii) leveraging the constraint satisfaction properties of CBFs to guarantee the fulfillment of the original task while minimizing an additional objective. The main advantages of the proposed CBF-PA approach are threefold. First, the introduction of a CBF with a relaxation variable addresses the potential empty intersection of optimal sets for the original and additional tasks. Second, the proposed algorithm operates as a closed-loop system, thereby reducing computational complexity. Finally, we validate the efficacy of the method through experiments on both low- and high-dimensional systems, demonstrating that CBF-PA outperforms traditional imitation and transfer learning methods by achieving lower additional costs while strictly maintaining the deviation from the optimal cost of the original task throughout the learning process.

One shortcoming of the proposed algorithm discussed in this paper is that standard RL algorithms often require the addition of exploration noise to the control policy for improved data exploration. However, this noise may negatively impact the performance of the proposed algorithm during training. Future work could focus on improving the definition of exploration noise within the context of the proposed algorithm. Additionally, the optimal critic and actor in the system may have nonconvex structures, potentially leading to multiple optimal solutions for the proposed methods.

\bibliographystyle{unsrt}
\bibliography{refs_hao, zihao, ref_zehui, ref_miguel}

\section{Appendix}
This appendix presents the proofs and experimental details underlying the results reported in the main text.
\subsection{Proof of Lemma~\ref{lemma1}}
To achieve the closed-form solution of the QP problem \eqref{eq_cbfqp}, we consider the following Lagrangian associated with \eqref{eq_cbfqp}: \begin{equation}
   \begin{aligned}
       \mathcal{L}(\boldsymbol{a}, c, \lambda_1, \lambda_2) = \frac{1}{2}\boldsymbol{a}'\boldsymbol{a} + \frac{w}{2} c^2 - \lambda_1 \hat{g}(\boldsymbol{a},c) - \lambda_2 c ,\nonumber
   \end{aligned}
   \end{equation} where $\lambda_1\geq 0$ and $\lambda_2\geq 0$ are Lagrange multipliers of function $\hat{g}(\boldsymbol{a},c)$ and $c$, respectively. By following the Karush-Kuhn-Tucker (KKT) condition, the following holds:
\begin{equation}\label{eq_KKT_0}
\begin{cases}
\frac{\partial \mathcal{L}}{\partial \boldsymbol{a}} = \boldsymbol{a}' - \lambda_1 L_g = \mathbf{0}_p,\\
\frac{\partial \mathcal{L}}{\partial c} = wc - \lambda_1 \frac{\partial \kappa(h(\boldsymbol{\theta})+c)}{\partial c}- \lambda_2 = 0,\\
-\lambda_1 (L_f  + L_g\boldsymbol{a} + \kappa(h(\boldsymbol{\theta})+c) = 0,\\
- \lambda_2 c= 0,\quad \lambda_1\geq 0,\quad \lambda_2\geq 0.
\end{cases} 
\end{equation}
Let $L_a = L_f  + \gamma_{\mathrm{h}}(G(\boldsymbol{\theta}_G^*) - G(\boldsymbol{\theta}))$, if $\kappa(h(\boldsymbol{\theta}) +c) = \gamma_{\mathrm{h}}(h(\boldsymbol{\theta})+c)$ with $\gamma_{\mathrm{h}}>0$, the closed-form solution of \eqref{eq_KKT_0} is:
\begin{equation}
    \begin{cases}
      \text{if}\ \lambda_1 = 0, \lambda_2 \geq 0,  \quad &\boldsymbol{a}(\boldsymbol{\theta}) = \mathbf{0}_p ,\  c^* = 0, \\ 
      \text{if}\ \lambda_1 > 0, \lambda_2 = 0, \quad  & \boldsymbol{a}(\boldsymbol{\theta}) = \frac{-L_aL_g'}{L_g L_g' + \gamma_{\mathrm{h}}^2/w} , c^* =  \frac{-\gamma_{\mathrm{h}}L_a}{w L_gL_g' +\gamma_{\mathrm{h}}^2},\\
      \text{if}\ \lambda_1 > 0, \lambda_2 > 0, \quad &\boldsymbol{a}(\boldsymbol{\theta}) = \frac{-L_a L_g'}{L_g L_g' } , c^* = 0.
    \end{cases} \nonumber
\end{equation} $\hfill \blacksquare$

\subsection{Proof of Lemma~\ref{lemma_cbf_valid}}
Because the closed-loop controller $\boldsymbol{a}(\boldsymbol{\theta})$ to the nominal gradient descent update is not bounded and can take any necessary magnitude and direction, it is sufficient to show that when $L_g \equiv \mathbf{0}_p'$, the CBF inequality constraint $L_f  + \kappa(h(\boldsymbol{\theta})+c) \geq 0$ holds \cite{ames2019control}. Following the parameter dynamics in \eqref{eq_dyn_theta}, $\forall c\geq 0$, the corresponding Lie derivatives are given by:
    \begin{subequations}
        \begin{align}
            L_f  &=-\nabla_{\boldsymbol{\theta}}h(\boldsymbol{\theta})'\nabla_{\boldsymbol{\theta}}J(\boldsymbol{\theta}),  \label{eq:lfh_proof}\\
            L_g &= \nabla_{\boldsymbol{\theta}}h(\boldsymbol{\theta})'. \label{eq:lgh_proof}
        \end{align}        
    \end{subequations}
Clearly, \eqref{eq:lgh_proof} can only be nullified if $\nabla_{\boldsymbol{\theta}}h(\boldsymbol{\theta}) = \mathbf{0}_p$, which would also nullify \eqref{eq:lfh_proof}. Therefore, when $L_g = \mathbf{0}_p'$, $L_f  + \kappa(h(\boldsymbol{\theta})+c) \geq 0$ becomes $\kappa(h(\boldsymbol{\theta})+c) \geq 0$, which always holds by the definition of class-$\mathcal{K}$ functions \cite{ames2019control}. $\hfill \blacksquare$

\subsection{Proof of Theorem~\ref{thm_a_star}}
Since the parameter is initialized as $\boldsymbol{\theta} = \boldsymbol{\theta}_G^*$, the constraints in \eqref{eq_presol_cons} holds for any $c^*\geq 0$, which implies that the initial $\boldsymbol{\theta}\in\mathcal{C}_{\theta, c^*}$. Furthermore, since $(\boldsymbol{a}(\boldsymbol{\theta}), c^*)$ in \eqref{eq_alg_qp_sol} represents solution satisfying the KKT condition of the QP problem in \eqref{eq_cbfqp}, and Lemma~\ref{lemma_cbf_valid} establishes that $h(\boldsymbol{\theta})+c^*$ in \eqref{eq_presol_cons} is a valid CBF, it follows from Definition~\ref{def2} that, under the closed-loop dynamics \eqref{eq_dyn_theta}, the controller $\boldsymbol{a}(\boldsymbol{\theta})$ satisfying \eqref{eq_cbf_constraint} guarantees the forward invariance of the constraint-admissible set $\mathcal{C}_{\theta, c^*}$. Therefore, we conclude that $\boldsymbol{\theta}\in\mathcal{C}_{\theta, c^*}$. $\hfill \blacksquare$

\end{document}